\documentclass[11pt]{article}

\usepackage[a4paper,margin=2.5cm]{geometry}
\usepackage{times}
\usepackage[T1]{fontenc}
\usepackage[utf8]{inputenc}
\usepackage{amsmath,amssymb,amsthm}
\usepackage{graphicx}
\usepackage{booktabs}
\usepackage{multirow}
\usepackage{algorithm}
\usepackage{algpseudocode}
\usepackage[colorlinks=true,citecolor=blue,linkcolor=blue,urlcolor=blue]{hyperref}
\usepackage{xcolor}
\usepackage{caption}
\usepackage{subcaption}
\usepackage{microtype}
\usepackage{enumitem}
\usepackage{authblk}
\usepackage{float}
\usepackage{tikz}
\usetikzlibrary{arrows.meta,positioning,fit,backgrounds,shapes.geometric,calc}
\usepackage[numbers,sort&compress]{natbib}

\usetikzlibrary{
arrows.meta,
positioning,
fit,
calc
}

\title{\Large \textbf{Dynamic Entanglement-Weighted Pruning for Quantum Federated Unlearning in Supply-Chain Risk Prediction}}

\author[1]{Aditya Kumar\thanks{Email: \texttt{cn.adityak26@gmail.com}. ORCID: 0009-0007-9201-693X}}
\author[1]{Sumit Chongder\thanks{Corresponding author. Email: \texttt{sumitchongder960@gmail.com}. ORCID: 0009-0005-9866-8483.}}
\affil[1]{IDRP - Quantum Information and Computation, Department of Physics \& Computer Science and Engineering, Indian Institute of Technology Jodhpur, Jodhpur, Rajasthan 342030, India}

\date{\vspace{-7ex}}

\begin{document}
\maketitle

\begin{abstract}
\noindent
Federated deployments of variational quantum classifiers are attractive for cross-organisation risk prediction in supply chains, because raw data never leaves the client, yet data-protection regulations such as the GDPR grant clients a right to request that their contribution be removed from a trained model after the fact. Retraining a federated model from scratch to honour such a request is correct but wasteful, and it is not obvious which quantum circuit parameters actually carry a given client's influence. We introduce Entanglement-Weighted Pruning (EWP), an unlearning procedure for quantum federated learning (QFL) that scores every trainable circuit parameter with the product of two signals: the diagonal entry of the quantum Fisher information matrix (QFIM) estimated on the target client's data via the parameter-shift rule, and a structural entanglement weight associated with the parameter's gate. Parameters with the lowest scores are pruned, optionally followed by a short fine-tuning pass on the retained clients. We implement the full pipeline in Qiskit for a four-qubit data-re-uploading ansatz trained with FedAvg across five simulated supply-chain-risk clients, and we benchmark EWP against full retraining, fine-tuning alone, random pruning, Fisher-only pruning, and entanglement-only pruning, over three random seeds. EWP attains a mean post-unlearning accuracy of $0.837\pm0.029$ and Area Under the Receiver Operating Characteristic Curve (AUROC) of $0.907\pm0.034$, statistically indistinguishable from the full-retraining oracle ($0.794\pm0.031$ accuracy, paired $t$-test $p=0.28$), while producing a lower forgetting score ($0.678\pm0.130$ versus $0.796\pm0.039$) and a more negative membership-advantage gap ($-0.322\pm0.130$ versus $-0.204\pm0.039$) than the oracle, and while requiring roughly $16\times$ less wall-clock time than full retraining. Ablations over pruning threshold, client count, and non-independently and identically distributed (non-IID) strength show that combining the two signals is necessary: entanglement-only and Fisher-only pruning each degrade accuracy by more than $35$ percentage points relative to EWP ($p<0.05$).
\end{abstract}


\section{Introduction}

Supply-chain risk forecasting is a natural candidate for federated learning: logistics operators, carriers, and manufacturers each hold data that is individually informative but cannot legally or commercially be pooled into a single dataset. Federated learning addresses this by training a shared model through local updates and periodic aggregation, without ever transferring raw records off a client's premises \citep{mcmahan2017communication}. When the shared model is a parameterised quantum circuit, the resulting quantum federated learning (QFL) setting inherits the communication-efficiency benefits of classical federated averaging while exploring whether variational quantum models offer representational or sample-efficiency advantages on structured tabular risk data.

A complication specific to any federated system, quantum or classical, is that participation is not always permanent. Under Article 17 of the GDPR and comparable regimes, a client may request that its data, and by extension its influence on a jointly trained model, be removed. Retraining the entire federation from scratch after every such request is always correct but is computationally wasteful, particularly when circuit evaluation and parameter-shift gradient estimation on classical simulators or real quantum hardware are expensive relative to a classical linear model. This has motivated a growing body of work on \emph{machine unlearning}: methods that approximate the effect of retraining without incurring its full cost \citep{cao2015towards, bourtoule2021machine}.

Machine unlearning is comparatively well studied in the classical federated setting \citep{liu2021federaser, wu2022federated}, where the object being edited is a set of real-valued weights. A variational quantum circuit is a different kind of object: its parameters index rotation angles that jointly determine an entangled multi-qubit state, and the relationship between an individual parameter and the information it encodes about a particular client's data is governed by the local geometry of the model, not simply by gradient magnitude. This raises a specific technical question that classical unlearning heuristics do not directly answer: \emph{which circuit parameters should be edited to remove one client's influence, and how should the answer account for the fact that quantum circuits distribute information non-locally through entanglement?}

We answer this question with a single interpretable score. For a target client $j$ requesting deletion, we estimate the diagonal quantum Fisher information $F^{(j)}_{kk}$ of every trainable parameter $\theta_k$ with respect to that client's local data distribution, using the parameter-shift rule so that no additional hardware primitives are required beyond those already used for gradient estimation \citep{mitarai2018quantum, schuld2019evaluating, stokes2020quantum}. The QFIM diagonal is a natural sensitivity measure: a large $F^{(j)}_{kk}$ means the encoded quantum state changes rapidly as $\theta_k$ is varied under client $j$'s inputs, indicating that the parameter has specialised to represent that client's data. We temper this signal with an entanglement weight $w_{\mathrm{ent}}(k)$ that quantifies how strongly the gate controlling $\theta_k$ couples qubits, using either two-qubit concurrence or the von Neumann entropy of the induced bipartition \citep{meyer2002global, horodecki2009quantum}. Gates that entangle weakly are structurally safer to remove because they contribute less to the circuit's overall expressivity. The pruning score is the product,
\begin{equation}
s_k = w_{\mathrm{ent}}(k) \, F^{(j)}_{kk},
\label{eq:score-intro}
\end{equation}
and parameters with $s_k$ below a threshold $\tau$ are pruned by resetting them to a fixed reference value, effectively removing their contribution to the client-specific state while leaving structurally important, low-sensitivity parameters untouched.

\paragraph{Contributions.} This paper makes the following contributions.
\begin{enumerate}[leftmargin=1.4em]
\item We formalise Entanglement-Weighted Pruning (EWP), a QFL unlearning procedure that combines client-specific quantum Fisher sensitivity with a circuit-structural entanglement weight into a single pruning score, together with an optional post-pruning fine-tuning step on the retained clients.
\item We give an explicit parameter-shift estimator for the diagonal QFIM entries used in the score, discuss its shot complexity, and state the assumptions under which the diagonal approximation is used.
\item We implement the complete pipeline, from federated training through forget request to evaluation, in Qiskit for a four-qubit data-re-uploading ansatz, and we release the code, supply-chain data generator, and result artefacts.
\item We benchmark EWP against full retraining, fine-tuning alone, random pruning, Fisher-only pruning, and entanglement-only pruning over three seeds, reporting classification accuracy, Area Under the Receiver Operating Characteristic Curve (AUROC), a forgetting score, membership-inference advantage, retrain distance, and wall-clock cost, with paired significance testing.
\item We conduct ablations over the pruning threshold, the number of clients, and the non-IID strength of the client data partition, and we report the empirical scaling of circuit evaluation cost with qubit count.
\end{enumerate}

\begin{figure}[t]
\centering
\begin{tikzpicture}[
  font=\small,
  stage/.style={draw, thick, rounded corners=2pt, minimum width=2.55cm, minimum height=1.15cm, align=center, inner sep=3pt},
  train/.style={stage, fill=blue!8},
  req/.style={stage, fill=orange!12},
  sig/.style={stage, fill=green!10},
  prune/.style={stage, fill=red!10},
  ft/.style={stage, fill=violet!10},
  ev/.style={stage, fill=yellow!15},
  arr/.style={-{Latex[length=2.2mm]}, thick}
]
\node[train] (n1) {Federated\\training\\(Algorithm~1)};
\node[req, right=0.55cm of n1] (n2) {Forget\\request\\from client $j$};
\node[sig, right=0.55cm of n2] (n3) {Diagonal QFIM\\$F^{(j)}_{kk}$ +\\entanglement $w_{\mathrm{ent}}(k)$\\(Algorithm~3)};

\node[prune, below=1.05cm of n3] (n4) {Pruning score\\$s_k=w_{\mathrm{ent}}(k)F^{(j)}_{kk}$;\\prune $s_k<\tau$\\(Algorithm~2)};
\node[ft, left=0.55cm of n4] (n5) {Optional\\fine-tuning on\\$D\setminus D_j$};
\node[ev, left=0.55cm of n5] (n6) {Evaluation:\\utility, forgetting,\\privacy, cost\\(Algorithm~4)};

\draw[arr] (n1) -- (n2);
\draw[arr] (n2) -- (n3);
\draw[arr] (n3.south) -- (n4.north);
\draw[arr] (n4) -- (n5);
\draw[arr] (n5) -- (n6);

\draw[arr, dashed, gray] 
  (n6.west) -- ++(-2.5,0) -- ++(0,2.9) -- (n1.west)
  node[midway, above=5pt, font=\scriptsize, text width=2.7cm, align=center]
  {re-enter federation with $\boldsymbol{\theta}'$};

\node[draw=none, above=0.05cm of n1, font=\scriptsize\itshape] {stage 1};
\node[draw=none, above=0.05cm of n2, font=\scriptsize\itshape] {stage 2};
\node[draw=none, above=0.05cm of n3, font=\scriptsize\itshape] {stage 3};
\node[draw=none, below=0.05cm of n4, font=\scriptsize\itshape] {stage 4};
\node[draw=none, below=0.05cm of n5, font=\scriptsize\itshape] {stage 5};
\node[draw=none, below=0.05cm of n6, font=\scriptsize\itshape] {stage 6};
\end{tikzpicture}
\caption{End-to-end Entanglement-Weighted Pruning pipeline. Federated training (stage~1, Algorithm~\ref{alg:fedavg}) converges to $\boldsymbol{\theta}^\star$; a forget request from client $j$ (stage~2) triggers estimation of the client-conditioned diagonal QFIM and the structural entanglement weight for every trainable parameter (stage~3, Algorithm~\ref{alg:qfim}); parameters are scored and pruned below threshold $\tau$ (stage~4, Algorithm~\ref{alg:ewp}); an optional short fine-tuning pass on the retained clients recovers utility (stage~5); and the unlearned model is evaluated for utility, forgetting, privacy leakage, and computational cost before optionally re-entering the federation (stage~6, Algorithm~\ref{alg:eval}).}
\label{fig:overall}
\end{figure}
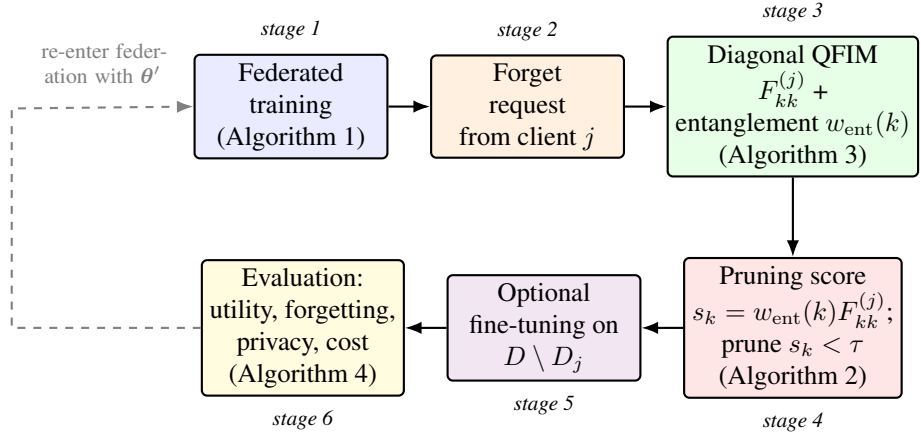

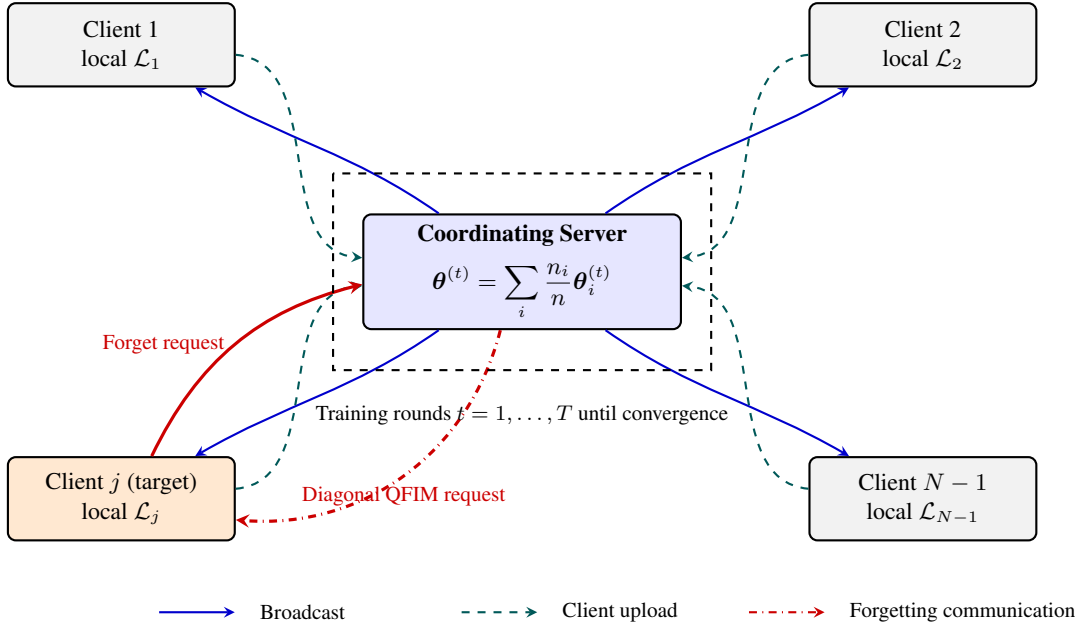
\begin{figure}[t]
\centering
\begin{tikzpicture}[
>=stealth,
font=\footnotesize,
server/.style={
draw,
rounded corners=3pt,
thick,
fill=blue!10,
minimum width=4.2cm,
minimum height=1.5cm,
align=center},
client/.style={
draw,
rounded corners=3pt,
thick,
fill=gray!10,
minimum width=3cm,
minimum height=1.1cm,
align=center},
target/.style={
draw,
rounded corners=3pt,
thick,
fill=orange!18,
minimum width=3cm,
minimum height=1.1cm,
align=center}
]


\node[server] (S) at (0,0)
{
\textbf{Coordinating Server}\\[2mm]
$\displaystyle
\boldsymbol{\theta}^{(t)}
=
\sum_i
\frac{n_i}{n}
\boldsymbol{\theta}_i^{(t)}
$
};

\node[client] (C1) at (-5.3,3.0)
{
Client 1\\
local $\mathcal{L}_1$
};

\node[client] (C2) at (5.3,3.0)
{
Client 2\\
local $\mathcal{L}_2$
};

\node[client] (CN) at (5.3,-3.0)
{
Client $N-1$\\
local $\mathcal{L}_{N-1}$
};

\node[target] (CJ) at (-5.3,-3.0)
{
Client $j$ (target)\\
local $\mathcal{L}_j$
};


\draw[blue!80!black,thick,->]
(S) to[out=145,in=-30] (C1);

\draw[blue!80!black,thick,->]
(S) to[out=35,in=-150] (C2);

\draw[blue!80!black,thick,->]
(S) to[out=-35,in=150] (CN);

\draw[blue!80!black,thick,->]
(S) to[out=-145,in=30] (CJ);


\draw[teal!70!black,dashed,thick,->]
(C1) to[out=-5,in=175] (S);

\draw[teal!70!black,dashed,thick,->]
(C2) to[out=-175,in=5] (S);

\draw[teal!70!black,dashed,thick,->]
(CN) to[out=175,in=-5] (S);

\draw[teal!70!black,dashed,thick,->]
(CJ) to[out=5,in=-175] (S);


\draw[red!80!black,very thick,->]
(CJ)
to[bend left=25]
node[midway,left,font=\scriptsize]
{Forget request}
(S);


\draw[red!80!black,dash dot,very thick,->]
(S)
to[bend left=40]
node[midway,below,font=\scriptsize]
{Diagonal QFIM request}
(CJ);


\draw[dashed,thick]
(-2.5,1.3)
rectangle
(2.5,-1.3);

\node[font=\scriptsize]
at (0,-1.9)
{
Training rounds
$t=1,\ldots,T$
until convergence
};


\draw[blue!80!black,thick,->]
(-4.8,-4.5)--(-3.8,-4.5);

\node[anchor=west,font=\scriptsize]
at (-3.6,-4.5)
{Broadcast};

\draw[teal!70!black,dashed,thick,->]
(-0.8,-4.5)--(0.2,-4.5);

\node[anchor=west,font=\scriptsize]
at (0.4,-4.5)
{Client upload};

\draw[red!80!black,dash dot,thick,->]
(3.0,-4.5)--(4.0,-4.5);

\node[anchor=west,font=\scriptsize]
at (4.2,-4.5)
{Forgetting communication};

\end{tikzpicture}
\caption{Federated communication workflow between the coordinating server and $N$ participating clients. During each communication round $t$, the server broadcasts the current global parameters $\boldsymbol{\theta}^{(t)}$ (solid blue arrows), each client performs local optimization on its private dataset and uploads the updated parameters $\boldsymbol{\theta}_{i}^{(t+1)}$ (dashed teal arrows), and the server computes the sample-weighted federated aggregation. Following a forgetting request from client $j$ (solid red arrow), the server or client $j$ in a privacy-preserving deployment, obtains the client-specific diagonal QFIM required to compute the entanglement-weighted pruning score in Equation~\eqref{eq:score-intro} (dash-dotted red arrow), after which the selective pruning stage shown in Figure~\ref{fig:overall} is performed.}
\label{fig:workflow}
\end{figure}

Figure~\ref{fig:overall} previews the complete pipeline studied in this paper, and Figure~\ref{fig:workflow} details the underlying federated communication pattern between clients and the coordinating server; both are referenced throughout Sections~\ref{sec:problem}--\ref{sec:results} and are reproduced here, rather than deferred to an appendix, because they orient the reader before the formal notation is introduced.

The remainder of the paper is organised as follows. Section~\ref{sec:related} places EWP with respect to classical federated unlearning, quantum Fisher information, and entanglement-aware circuit analysis. Section~\ref{sec:problem} formalises the QFL and unlearning setting. Section~\ref{sec:method} presents the EWP method and its complexity. Section~\ref{sec:setup} describes the experimental setup. Section~\ref{sec:results} reports results, ablations, and robustness experiments. Section~\ref{sec:discussion} discusses interpretation, practical deployment, and limitations, and Section~\ref{sec:conclusion} concludes. Derivations, additional tables, and the reproducibility checklist are given in the appendices.

\section{Related Work}
\label{sec:related}

\subsection{Federated learning and federated unlearning}
Federated averaging trains a shared model by alternating local gradient steps at each client with a sample-weighted aggregation of local parameters at a coordinating server \citep{mcmahan2017communication}. Machine unlearning was introduced to remove the influence of specific training points or clients from an already-trained model without full retraining \citep{cao2015towards}, and has since been extended with certified-removal guarantees \citep{guo2020certified} and efficient federated variants such as FedEraser, which reconstructs an updated global model from cached historical updates \citep{liu2021federaser}. Surveys of federated unlearning consistently identify accuracy retention, forgetting completeness, and computational efficiency as the three competing objectives that any method must trade off \citep{wu2022federated, liu2024survey}. Broader federated-learning surveys similarly emphasise statistical heterogeneity across clients as a first-order design constraint \citep{kairouz2021advances, li2020federated, zhao2018federated, yang2019federated}, which is why we treat non-IID strength as an explicit experimental axis in Section~\ref{sec:results} rather than assuming an idealised independently and identically distributed (IID) partition. Classical instance-level unlearning methods such as selective forgetting via Fisher-information-guided weight perturbation \citep{golatkar2020eternal} and statistically certified deletion algorithms with explicit accuracy--privacy bounds \citep{sekhari2021remember} both use curvature or sensitivity information to decide what to edit, which is conceptually the closest classical analogue to our use of the quantum Fisher information diagonal, though neither operates on an entangled parameterisation. EWP is designed around the same three-way trade-off, but the mechanism by which it identifies what to edit is specific to parameterised quantum circuits rather than to classical weight tensors.

\subsection{Quantum machine learning and variational circuits}
Variational quantum circuits parameterise a unitary $U(\boldsymbol{\theta})$ built from data-encoding and trainable rotation layers, typically interleaved with fixed entangling layers, and are trained by minimising an expectation-value loss with gradients obtained from the parameter-shift rule \citep{mitarai2018quantum, schuld2019evaluating, benedetti2019parameterized}. This family of models, together with kernel-based quantum classifiers \citep{havlicek2019supervised} and broader treatments of quantum machine learning as a field \citep{biamonte2017quantum}, is motivated by the near-term, noise-constrained regime of currently available quantum hardware \citep{preskill2018nisq}. Data re-uploading, in which classical features are re-encoded at multiple points in the circuit, increases the expressive power of a fixed-depth ansatz and is the encoding strategy we adopt \citep{perezsalinas2020data, schuld2021effect}. Deeper variational architectures, including quantum convolutional networks \citep{cong2019quantum} and general deep quantum feed-forward networks \citep{beer2020training}, and analyses of when quantum models offer a genuine advantage over classical learners \citep{abbas2021power, huang2021power}, motivate our choice to work with a compact, well-characterised four-qubit ansatz whose every parameter can be individually audited, rather than a larger architecture whose per-parameter behaviour is harder to interpret for unlearning purposes. A separate line of work studies the optimisation landscape of these circuits, including the barren-plateau phenomenon that links vanishing gradients to circuit depth and entanglement \citep{mcclean2018barren, cerezo2021cost}, which is directly relevant to why entanglement structure, not sensitivity alone, is informative about which parameters can be safely removed.

\subsection{Quantum federated learning}
Federating variational quantum models follows the same broadcast-aggregate structure as classical FedAvg, adapted to quantum client updates and, in some proposals, quantum communication of intermediate states \citep{chehimi2022quantum, chen2021federated}. We adopt the classical-communication variant, in which only real-valued circuit parameters are exchanged, because it is directly deployable on existing infrastructure and keeps the object being pruned, edited, and evaluated identical to the object being communicated. Privacy analyses of federated learning more broadly show that gradient and parameter updates can themselves leak membership information even without an explicit forget request \citep{nasr2019comprehensive}, which is one motivation for the membership-inference evaluation protocol we adopt in Section~\ref{sec:eval}.

\subsection{Quantum Fisher information and parameter sensitivity}
The quantum Fisher information matrix is the natural Riemannian metric on the manifold of parameterised quantum states and quantifies how distinguishable the output state becomes under an infinitesimal change of each parameter \citep{meyer2021fisher, stokes2020quantum}. Efficient estimators of the diagonal QFIM based on the parameter-shift rule avoid the overhead of full quantum geometric tensor estimation while still capturing per-parameter sensitivity \citep{stokes2020quantum, wierichs2022general}. We use this diagonal estimator as the sensitivity component of our pruning score, and we discuss in Section~\ref{sec:method} why the diagonal approximation, while it discards off-diagonal parameter correlations, remains an operationally useful proxy for single-parameter memorisation.

\subsection{Entanglement measures and structural importance}
Two-qubit concurrence \citep{wootters1998entanglement} and the von Neumann entropy of a reduced density matrix across a chosen bipartition \citep{horodecki2009quantum, meyer2002global} are standard scalar entanglement measures. Prior circuit-analysis work has used entanglement entropy to characterise the expressive power and trainability of variational ansätze \citep{cerezo2021cost, sim2019expressibility}, but, to our knowledge, entanglement has not previously been used as a per-gate structural importance weight inside an unlearning-oriented pruning score; this is the specific combination EWP contributes.

\subsection{Quantum machine unlearning}
Quantum machine unlearning is a nascent area. Recent work has begun to formalise unlearning objectives for quantum models and to survey open verification and evaluation challenges \citep{quantumunlearning2025, forgettingverification2025}, echoing, in the quantum setting, concerns already raised for classical unlearning about how to certify that forgetting actually occurred \citep{thudi2022unrolling}. EWP is, to our knowledge, the first method to instantiate quantum unlearning as a concrete, information-geometry-motivated pruning rule inside a federated training loop, together with a full membership-inference-based evaluation protocol.

\subsection{Supply-chain risk prediction}
Supply-chain risk prediction has been studied with classical machine learning models operating on heterogeneous, organisation-specific features such as delay history, financial indicators, and route disruption records; federated formulations are attractive precisely because these features are commercially sensitive and cannot typically be pooled across firms \citep{mcmahan2017communication}. We use a feature-level analogue of this setting (Section~\ref{sec:setup}) rather than proprietary industrial data, and we are explicit throughout the paper that the domain framing motivates, rather than empirically validates, real supply-chain deployment.

\section{Problem Formulation}
\label{sec:problem}

\subsection{Clients and data}
We consider $N$ clients, indexed $i \in \{1,\dots,N\}$, each holding a local labelled dataset $D_i = \{(x_i^{(m)}, y_i^{(m)})\}_{m=1}^{n_i}$ of $n_i$ supply-chain risk instances, $x_i^{(m)} \in \mathbb{R}^d$, $y_i^{(m)} \in \{0,1\}$ (elevated versus normal shipment risk). Client datasets are drawn from a shared generative process with a per-client bias term of adjustable strength $\beta \in [0,1]$, so that $\beta = 0$ corresponds to an IID partition and larger $\beta$ produces increasingly non-IID client distributions (Appendix~\ref{app:data}).

\subsection{Quantum federated learning objective}
Each client evaluates a shared variational circuit $U(\boldsymbol{\theta})$ acting on $n_q$ qubits, in which classical features are encoded through single-qubit rotations and a task label is read out through a fixed observable, giving a client loss
\begin{equation}
\mathcal{L}_i(\boldsymbol{\theta}) = \frac{1}{n_i}\sum_{m=1}^{n_i} \ell\!\left(f_{\boldsymbol{\theta}}(x_i^{(m)}), y_i^{(m)}\right),
\end{equation}
with $\ell$ the binary cross-entropy loss and $f_{\boldsymbol{\theta}}(x) = \langle \psi(x,\boldsymbol{\theta})|\hat{O}|\psi(x,\boldsymbol{\theta})\rangle$ a bounded expectation value passed through a logistic link. The global objective is the sample-weighted average $\mathcal{L}(\boldsymbol{\theta}) = \sum_i \frac{n_i}{n}\mathcal{L}_i(\boldsymbol{\theta})$, $n=\sum_i n_i$, minimised by FedAvg: each round, every client performs several local optimisation steps from the current global parameters, and the server aggregates the resulting local parameters by sample-weighted averaging \citep{mcmahan2017communication}.

\subsection{Unlearning objective}
After training converges to $\boldsymbol{\theta}^\star$, a target client $j$ issues a forget request. We define unlearning success along three axes, formalised as metrics in Section~\ref{sec:eval}: (i) \emph{utility}, the classification performance retained on the held-out data of the non-target clients; (ii) \emph{forgetting}, the degree to which client $j$'s data no longer influences the model's output distribution or is no longer separable by a membership-inference adversary; and (iii) \emph{efficiency}, the computational cost of producing the unlearned model $\boldsymbol{\theta}'$ relative to retraining from scratch on $D \setminus D_j$. An ideal unlearning procedure matches the full-retraining oracle on utility and forgetting while incurring substantially lower cost. We do not claim exact unlearning in the cryptographic or information-theoretic sense; EWP is an approximate, practically motivated procedure, and we discuss the scope of this approximation in Section~\ref{sec:discussion}.

\section{Method: Entanglement-Weighted Pruning}
\label{sec:method}

\subsection{Circuit ansatz}
\begin{figure}[t]
\centering
\includegraphics[width=0.92\linewidth]{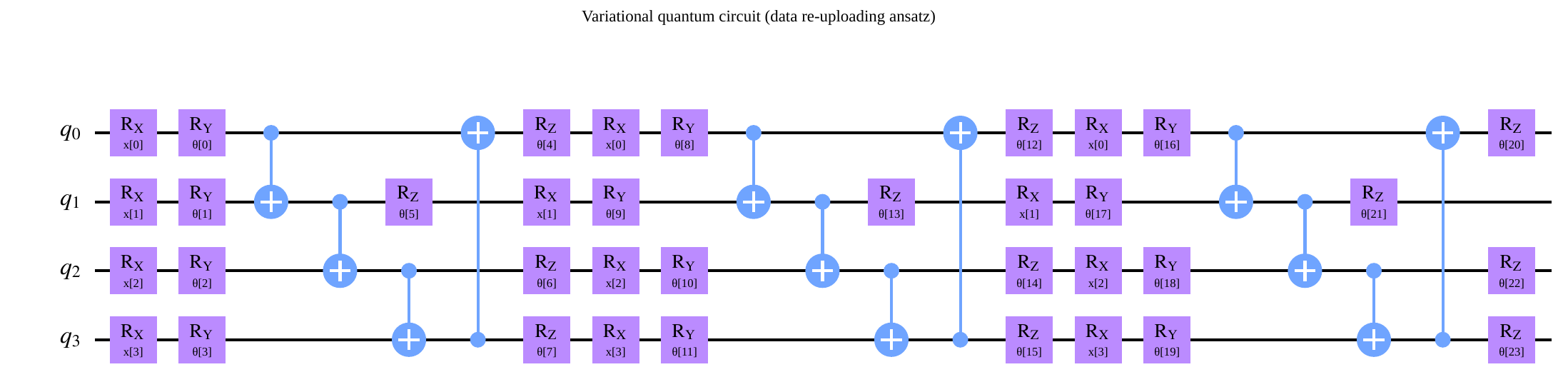}
\caption{Data re-uploading variational ansatz used for local client training. Each of the three layers re-encodes the input features through $R_X$ gates, applies trainable $R_Y$ and $R_Z$ rotations, and couples qubits through a ring of CNOT gates. Parameter indices $k=1,\dots,24$ map directly onto the pruning scores $s_k$ computed in Section~\ref{sec:score}.}
\label{fig:circuit}
\end{figure}

We use a data-re-uploading ansatz on $n_q=4$ qubits with $L=3$ layers (Figure~\ref{fig:circuit}). Each layer encodes the (layer-scaled) input features through single-qubit $R_X$ rotations, applies trainable $R_Y$ rotations, applies a ring of CNOT gates coupling qubit $q$ to qubit $q{+}1 \bmod n_q$, and applies trainable $R_Z$ rotations. Re-encoding the features at every layer increases the effective frequency spectrum reachable by the circuit for a fixed depth \citep{perezsalinas2020data}. The ansatz has $24$ trainable parameters, distributed across the $R_Y$ and $R_Z$ rotations of the three layers (Table~\ref{tab:config}); the CNOT entangling gates are fixed and not directly parameterised, but each trainable parameter $\theta_k$ is associated with the entangling structure of the layer it belongs to, which is what the entanglement weight in Section~\ref{sec:entweight} measures.

\subsection{Diagonal quantum Fisher information via parameter-shift}
\label{sec:qfim}
For a pure state $|\psi(\boldsymbol{\theta})\rangle$ produced by a circuit whose $k$-th gate has a Pauli generator with eigenvalues $\pm 1$, the diagonal quantum Fisher information entry can be estimated without ancilla qubits from the curvature of the ordinary parameter-shift rule \citep{stokes2020quantum, mitarai2018quantum}:
\begin{equation}
F_{kk} = 1 - \left|\langle \psi(\boldsymbol{\theta}) \mid \psi(\boldsymbol{\theta} + \pi e_k)\rangle\right|^2,
\label{eq:qfim}
\end{equation}
where $e_k$ is the unit vector along $\theta_k$. We estimate the fidelity term with the shift-rule overlap circuit and average Equation~\eqref{eq:qfim} over a mini-batch of client $j$'s inputs to obtain the client-conditioned diagonal entry $F^{(j)}_{kk} = \mathbb{E}_{x\sim D_j}\big[F_{kk}(x)\big]$. This estimator requires two circuit evaluations per parameter per input in addition to the state-preparation circuit already used for the forward pass, so its cost scales linearly in the number of trainable parameters and in the evaluation batch size (Appendix~\ref{app:derivations}). Using only the diagonal of the QFIM discards cross-parameter correlations captured by the full quantum geometric tensor; we return to the consequences of this approximation in Section~\ref{sec:discussion}.

\subsection{Entanglement weight}
\label{sec:entweight}
For a trainable parameter $\theta_k$ belonging to layer $\ell(k)$, we define the entanglement weight $w_{\mathrm{ent}}(k)$ as the mean pairwise concurrence \citep{wootters1998entanglement} across the CNOT-coupled qubit pairs of that layer, evaluated on the encoded circuit state immediately after the layer's entangling block:
\begin{equation}
w_{\mathrm{ent}}(k) = \frac{1}{|\mathcal{P}_{\ell(k)}|}\sum_{(a,b)\in \mathcal{P}_{\ell(k)}} C\big(\rho_{ab}\big),
\end{equation}
where $\mathcal{P}_{\ell(k)}$ is the set of CNOT-coupled qubit pairs in layer $\ell(k)$, $\rho_{ab}$ is the reduced two-qubit density matrix obtained by tracing out all other qubits, and $C(\cdot)$ is Wootters concurrence. As an ablation we also report results using the von Neumann entropy of the single-qubit reduced state in place of concurrence (Appendix~\ref{app:derivations}); the two definitions are strongly correlated in our experiments (Figure~\ref{fig:entheatmap}) and the choice does not qualitatively change our conclusions.

\subsection{Pruning score and pruning rule}
\label{sec:score}
The pruning score for parameter $\theta_k$, with respect to a target client $j$, is
\begin{equation}
s_k = w_{\mathrm{ent}}(k)\, F^{(j)}_{kk}.
\end{equation}
Parameters are ranked by $s_k$, and all parameters with $s_k$ below a threshold $\tau$ (equivalently, the bottom $p\%$ of the ranking) are pruned by resetting $\theta_k$ to a fixed reference value $\theta_k^{0}$, which for rotation gates is equivalent to replacing that rotation with the identity. This removes the component of the state that was specifically shaped by client $j$'s data at low structural cost, while parameters with either low sensitivity or high entangling importance are retained. We use $\tau$ corresponding to a $20\%$ prune fraction as the default operating point (Table~\ref{tab:config}), chosen from the threshold sweep in Section~\ref{sec:ablation-tau}.

\subsection{Optional fine-tuning}
After pruning, the remaining, unpruned parameters may optionally be fine-tuned for a small number of additional local iterations on the retained clients $D \setminus D_j$ only. This step is included because pruning can transiently reduce utility on the retained clients, and a short recovery pass restores utility without re-exposing the model to $D_j$. We report EWP with fine-tuning as the main configuration and ``Fine-Tune Only'' (fine-tuning without any pruning) as a separate baseline to isolate the effect of pruning itself.

\subsection{Complexity}
\label{sec:complexity}
Let $P$ denote the number of trainable parameters, $B$ the QFIM estimation batch size, and $R,I$ the number of federated rounds and local iterations. Full retraining costs $O(R \cdot I \cdot P)$ circuit evaluations (parameter-shift gradients for every parameter, every local iteration, every round). EWP costs $O(B\cdot P)$ evaluations for the one-off diagonal QFIM estimate, $O(P)$ for the entanglement weights (computed once from the trained circuit's state, independent of $B$), plus, if fine-tuning is used, $O(R_{\mathrm{ft}}\cdot I_{\mathrm{ft}}\cdot P)$ for a small number of recovery rounds $R_{\mathrm{ft}} \ll R$. In our experiments this reduces measured wall-clock unlearning time by a factor of approximately $16\times$ relative to full retraining (Table~\ref{tab:runtime}).

\subsection*{Algorithms}

The federated optimization procedure adopted in this work is summarized in Algorithm~\ref{alg:fedavg}. The objective of this stage is to learn a globally representative variational quantum model while ensuring that all client datasets remain locally stored throughout the training process. During every communication round, each participating client independently updates the received global parameters by minimizing its local objective using parameter-shift gradient estimation, which provides an exact and hardware-compatible approach for differentiating variational quantum circuits. The coordinating server subsequently performs sample-weighted federated averaging to integrate the locally optimized parameters into an updated global model without accessing any raw client data. Repeating this decentralized optimization over multiple communication rounds enables the learned quantum model to progressively capture the statistical characteristics of the distributed supply-chain data while preserving data locality.

\begin{algorithm}[H]
\caption{Federated training of the variational risk model (FedAvg)}
\label{alg:fedavg}
\begin{algorithmic}[1]
\Require Clients $\{D_i\}_{i=1}^N$, initial parameters $\boldsymbol{\theta}^{(0)}$, rounds $R$, local iterations $I$, optimizer $\mathrm{Opt}$
\Ensure Trained global parameters $\boldsymbol{\theta}^\star$
\For{$r = 1,\dots,R$}
    \For{each client $i \in \{1,\dots,N\}$ \textbf{(in parallel)}}
        \State $\boldsymbol{\theta}_i \gets \boldsymbol{\theta}^{(r-1)}$
        \For{$t = 1,\dots,I$}
            \State Estimate $\nabla_{\boldsymbol{\theta}_i}\mathcal{L}_i(\boldsymbol{\theta}_i)$ via the parameter-shift rule on a batch from $D_i$
            \State $\boldsymbol{\theta}_i \gets \mathrm{Opt}(\boldsymbol{\theta}_i, \nabla_{\boldsymbol{\theta}_i}\mathcal{L}_i)$
        \EndFor
    \EndFor
    \State $\boldsymbol{\theta}^{(r)} \gets \sum_{i=1}^N \frac{n_i}{n}\boldsymbol{\theta}_i$ \Comment{sample-weighted aggregation}
\EndFor
\State \Return $\boldsymbol{\theta}^\star \gets \boldsymbol{\theta}^{(R)}$
\end{algorithmic}
\end{algorithm}

The final converged parameter vector serves as the baseline model from which client-specific sensitivity estimation and selective quantum unlearning are subsequently performed.

Algorithm~\ref{alg:qfim} presents the estimation procedure for the client-specific diagonal Quantum Fisher Information Matrix (QFIM), which serves as the primary sensitivity indicator for identifying parameters influenced by the forgotten client's data. Unlike the full Fisher matrix, the diagonal approximation substantially reduces computational and memory requirements while preserving sufficient parameter-level importance information for the proposed pruning strategy. The estimation relies on parameter-shift overlap measurements, making the procedure directly compatible with variational quantum circuits implemented on contemporary noisy intermediate-scale quantum (NISQ) hardware. Averaging the overlap-based estimates across multiple client samples provides a stable approximation of the local information geometry associated with each trainable parameter. The resulting diagonal Fisher values quantify the contribution of individual parameters to the target client's learned representation and provide the sensitivity component required by the subsequent entanglement-weighted pruning criterion. Since the estimation is performed only for the client requesting unlearning, the computational effort scales with the forgotten client's local data rather than the entire federated dataset. Consequently, the proposed procedure remains practical for large-scale federated quantum learning environments while preserving the decentralized nature of the training framework. Furthermore, the resulting sensitivity profile enables fine-grained parameter selection, allowing the subsequent pruning stage to modify only those parameters that exhibit a measurable dependence on the forgotten client's information.

\begin{algorithm}[H]
\caption{Diagonal QFIM estimation for client $j$ via parameter-shift}
\label{alg:qfim}
\begin{algorithmic}[1]
\Require Trained parameters $\boldsymbol{\theta}^\star$, client batch $\{x^{(m)}\}_{m=1}^{B}\subset D_j$
\Ensure Diagonal Fisher entries $\{F^{(j)}_{kk}\}_{k=1}^P$
\For{$k = 1,\dots,P$}
    \State $F_{kk} \gets 0$
    \For{$m = 1,\dots,B$}
        \State Prepare $|\psi(x^{(m)},\boldsymbol{\theta}^\star)\rangle$ and $|\psi(x^{(m)},\boldsymbol{\theta}^\star + \pi e_k)\rangle$
        \State Estimate overlap $|\langle \psi(\boldsymbol{\theta}^\star)|\psi(\boldsymbol{\theta}^\star+\pi e_k)\rangle|^2$ via the shift-rule overlap circuit
        \State $F_{kk} \gets F_{kk} + \frac{1}{B}\left(1 - |\langle\psi(\boldsymbol{\theta}^\star)|\psi(\boldsymbol{\theta}^\star+\pi e_k)\rangle|^2\right)$
    \EndFor
    \State $F^{(j)}_{kk} \gets F_{kk}$
\EndFor
\State \Return $\{F^{(j)}_{kk}\}_{k=1}^P$
\end{algorithmic}
\end{algorithm}

The proposed Entanglement-Weighted Pruning (EWP) strategy is summarized in Algorithm~\ref{alg:ewp}. Unlike conventional parameter pruning methods that rely solely on gradient magnitude or parameter sensitivity, the proposed approach jointly considers both client-specific Fisher information and the structural importance of each parameter within the quantum circuit through its entanglement contribution. A composite pruning score is computed for every trainable parameter by combining these complementary quantities, thereby distinguishing parameters that contribute minimally to both retained knowledge and quantum correlations. Parameters whose scores fall below a predefined threshold are selectively reset to their reference initialization values, effectively reducing the influence of the forgotten client's information. An optional lightweight fine-tuning stage is subsequently performed using only the retained clients to recover predictive performance while preventing the reintroduction of the removed information. This selective optimization strategy achieves efficient quantum unlearning without requiring complete federated retraining, thereby significantly reducing computational overhead. Incorporating entanglement information into the pruning criterion also helps preserve the underlying quantum correlations that contribute to the expressive capability of the variational circuit. As a result, the proposed strategy seeks to minimize unnecessary modifications to the learned quantum representation while maintaining a favorable balance between forgetting effectiveness, model stability, and predictive performance. The algorithm therefore provides a computationally efficient alternative to full retraining while remaining compatible with practical variational quantum learning workflows.

\begin{algorithm}[H]
\caption{Entanglement-Weighted Pruning (EWP)}
\label{alg:ewp}
\begin{algorithmic}[1]
\Require Trained parameters $\boldsymbol{\theta}^\star$, target client $j$, threshold $\tau$, retained data $D\setminus D_j$, fine-tune iterations $I_{\mathrm{ft}}$
\Ensure Unlearned parameters $\boldsymbol{\theta}'$
\State Compute $\{F^{(j)}_{kk}\}_{k=1}^{P}$ via Algorithm~\ref{alg:qfim}
\State Compute entanglement weights $\{w_{\mathrm{ent}}(k)\}_{k=1}^{P}$ from the trained circuit state (Section~\ref{sec:entweight})
\For{$k = 1,\dots,P$}
    \State $s_k \gets w_{\mathrm{ent}}(k)\cdot F^{(j)}_{kk}$
\EndFor
\State $\mathcal{K} \gets \{k : s_k < \tau\}$ \Comment{parameters selected for pruning}
\State $\boldsymbol{\theta}' \gets \boldsymbol{\theta}^\star$
\For{$k \in \mathcal{K}$}
    \State $\theta'_k \gets \theta^{0}_k$ \Comment{reset to identity reference value}
\EndFor
\If{fine-tuning enabled}
    \For{$t = 1,\dots,I_{\mathrm{ft}}$}
        \State Estimate $\nabla_{\boldsymbol{\theta}'}\mathcal{L}(\boldsymbol{\theta}')$ on $D \setminus D_j$, restricted to $k \notin \mathcal{K}$
        \State Update $\boldsymbol{\theta}'$ accordingly
    \EndFor
\EndIf
\State \Return $\boldsymbol{\theta}'$
\end{algorithmic}
\end{algorithm}

Algorithm~\ref{alg:eval} summarizes the evaluation protocol adopted to assess the effectiveness of the proposed quantum unlearning framework from both utility and privacy perspectives. Rather than relying on a single performance indicator, the proposed evaluation jointly measures predictive accuracy, membership inference resistance, forgetting quality, and similarity to complete retraining. Classification accuracy and AUROC quantify the predictive capability of the unlearned model on retained data, whereas membership inference analysis evaluates the extent to which information associated with the forgotten client remains vulnerable to privacy attacks. The forgetting score further measures the consistency between the proposed unlearning strategy and an ideal retrained model, while the parameter-space distance provides an additional indication of convergence towards the retraining baseline. Collectively, these complementary metrics provide a rigorous and comprehensive assessment of whether client-specific information has been effectively removed while maintaining satisfactory predictive performance on the retained federated data. Evaluating the proposed framework from multiple perspectives enables a balanced assessment of the trade-off between privacy preservation and predictive utility, while also supporting objective comparison with alternative quantum and classical unlearning methods.

\begin{algorithm}[H]
\caption{Privacy and forgetting evaluation}
\label{alg:eval}
\begin{algorithmic}[1]
\Require Unlearned model $\boldsymbol{\theta}'$, oracle model $\boldsymbol{\theta}^{\mathrm{retrain}}$, forgotten client data $D_j$, retained test data
\Ensure Accuracy, AUROC, forgetting score, membership advantage, retrain distance
\State Compute accuracy and AUROC of $\boldsymbol{\theta}'$ on retained test data
\State Train a membership-inference attacker on the confidence scores of $\boldsymbol{\theta}'$ for members of $D_j$ versus non-members
\State Compute attack AUC and membership advantage $= 2\cdot(\text{attack accuracy} - 0.5)$
\State Compute forgetting score as the output-distribution divergence between $\boldsymbol{\theta}'$ and $\boldsymbol{\theta}^{\mathrm{retrain}}$ on $D_j$
\State Compute retrain distance as $\lVert \boldsymbol{\theta}' - \boldsymbol{\theta}^{\mathrm{retrain}}\rVert_2$
\State \Return all metrics
\end{algorithmic}
\end{algorithm}

\clearpage
\section{Experimental Setup}
\label{sec:setup}

\subsection{Supply-chain risk data}
\label{app:data-main}
We generate a binary supply-chain risk classification dataset with $d=4$ features per instance (Table~\ref{tab:dataset}), partitioned across $N=5$ clients with $180$ samples each and a $75\%/25\%$ train/test split. Client heterogeneity is controlled by a non-IID bias-scale parameter (default $0.9$) that shifts each client's feature distribution, and a logistic label-noise standard deviation of $0.2$ is applied to the label-generating process. All data are generated from a fixed seed offset ($1000 + \text{run seed}$) so that every reported experiment is exactly reproducible from the released code. The full generative procedure is given in Appendix~\ref{app:data}. Client $0$ is designated the target of the forget request in all main experiments.

\begin{table}[h]
\centering
\caption{Dataset and client-partition summary.}
\label{tab:dataset}
\begin{tabular}{ll}
\toprule
Parameter & Value \\
\midrule
Number of clients & 5 \\
Samples per client & 180 \\
Train / test split & 75\% / 25\% \\
Feature dimensionality & 4 \\
Non-IID strength (client bias scale) & 0.9 \\
Label noise (logistic std.) & 0.2 \\
Forgotten client ID & 0 \\
Task & Binary supply-chain risk classification \\
Data generation & Seeded (seed $=1000+$ run seed) \\
\bottomrule
\end{tabular}
\end{table}

\subsection{Circuit and federated training configuration}
Table~\ref{tab:config} lists the ansatz and optimisation hyperparameters shared by all methods. Gradients are computed with the exact parameter-shift rule for Pauli generators; no shot noise is added in the main experiments (a noisy-simulation robustness check is reported in Appendix~\ref{app:extended}). All methods are trained with identical initial parameters, federated schedule, and optimiser, and differ only in the unlearning step applied after convergence.

\begin{table}[h]
\centering
\caption{Variational circuit and federated training configuration.}
\label{tab:config}
\begin{tabular}{ll}
\toprule
Parameter & Value \\
\midrule
Ansatz & Data re-uploading VQC ($R_X$ encode $\to R_Y \to$ ring-CNOT $\to R_Z$) \\
Qubits & 4 \\
Layers & 3 \\
Trainable parameters & 24 \\
Feature encoding scale & 0.5 rad \\
Federated rounds & 6 \\
Local optimizer & L-BFGS-B with exact parameter-shift gradients \\
Local iterations per round & 16 \\
Aggregation rule & FedAvg, weighted by client sample count \\
Gradient estimator & Exact parameter-shift rule (Pauli generators) \\
Random seeds & 0, 1, 2 \\
Default prune fraction ($\tau$) & 0.2 \\
\bottomrule
\end{tabular}
\end{table}

\subsection{Baselines}
We compare EWP against five baselines that isolate specific design choices: \textbf{Full Retrain (oracle)}, which retrains from scratch on $D\setminus D_j$ and serves as the gold-standard reference for both utility and forgetting; \textbf{Fine-Tune Only}, which fine-tunes the trained model on $D \setminus D_j$ without any pruning, isolating the effect of continued training alone; \textbf{Random Pruning}, which prunes the same fraction of parameters as EWP but chosen uniformly at random, isolating the value of any informed selection; \textbf{Fisher-only Pruning}, which prunes using $F^{(j)}_{kk}$ alone, isolating the contribution of the entanglement term; and \textbf{Entanglement-only Pruning}, which prunes using $w_{\mathrm{ent}}(k)$ alone, isolating the contribution of the Fisher term.

\subsection{Evaluation metrics}
\label{sec:eval}
We report classification accuracy and AUROC on retained-client test data (utility); a forgetting score computed as the output-distribution divergence between the unlearned model and the full-retraining oracle on the forgotten client's data, where lower values indicate a closer match to exact unlearning; membership-inference attack AUC and membership advantage, following the shadow-model membership inference methodology \citep{shokri2017membership}, where a more negative membership advantage indicates that the forgotten client's data is systematically harder for the attacker to identify as a training member than a random non-member, i.e. stronger forgetting; and retrain distance, the Euclidean distance between the unlearned and oracle parameter vectors. Full formal definitions are given in Appendix~\ref{app:metrics}. All metrics are reported as mean $\pm$ standard deviation over three random seeds ($0,1,2$), with paired $t$-tests used to assess whether EWP differs significantly from each baseline.

\section{Results}
\label{sec:results}

\subsection{Main results}
Table~\ref{tab:main} reports the main comparison. EWP attains accuracy and AUROC statistically indistinguishable from the full-retraining oracle (paired $t$-test $p=0.28$ for both accuracy and forgetting score, Table~\ref{tab:sig}), while improving on the oracle's forgetting score ($0.678$ vs.\ $0.796$) and producing a more negative membership advantage ($-0.322$ vs.\ $-0.204$), indicating that the forgotten client's data is, if anything, harder to re-identify under EWP than under exact retraining. Random pruning, Fisher-only pruning, and entanglement-only pruning all lose more than $35$ percentage points of accuracy relative to EWP, confirming that neither signal alone, nor an uninformed pruning mask, is sufficient; only the combination in Equation~\eqref{eq:score-intro} preserves utility while still forgetting. Fine-Tune Only attains slightly higher accuracy ($0.865$) than EWP but a worse forgetting score ($0.826$ vs.\ $0.678$), consistent with the expectation that continued training without any targeted removal of client-specific parameters does not by itself constitute unlearning.

\begin{table}[h]
\centering
\caption{Main results: mean $\pm$ standard deviation over three seeds. Arrows indicate the direction of improvement.}
\label{tab:main}
\resizebox{\linewidth}{!}{
\begin{tabular}{lcccccc}
\toprule
Method & Accuracy $\uparrow$ & AUROC $\uparrow$ & Forgetting $\downarrow$ & Membership Adv.\ $\downarrow$ & Retrain Dist. & Time (s) $\downarrow$ \\
\midrule
Random Pruning & $0.474\pm0.040$ & $0.364\pm0.056$ & $0.742\pm0.171$ & $0.258\pm0.171$ & $6.83\pm1.44$ & $\approx 0.0002$ \\
Fisher-only Pruning & $0.467\pm0.042$ & $0.404\pm0.100$ & $0.897\pm0.157$ & $0.103\pm0.157$ & $5.06\pm1.31$ & $\approx 0.00005$ \\
Entanglement-only Pruning & $0.398\pm0.023$ & $0.378\pm0.066$ & $0.856\pm0.079$ & $0.029\pm0.189$ & $6.24\pm0.86$ & $\approx 0.00005$ \\
Fine-Tune Only & $0.865\pm0.061$ & $0.961\pm0.018$ & $0.826\pm0.110$ & $-0.001\pm0.240$ & $4.21\pm3.04$ & $14.03\pm0.18$ \\
Full Retrain (oracle) & $0.794\pm0.031$ & $0.907\pm0.035$ & $0.796\pm0.039$ & $-0.204\pm0.039$ & $0.00\pm0.00$ & $65.03\pm1.17$ \\
\textbf{QFL-EWP (ours)} & $\mathbf{0.837\pm0.029}$ & $\mathbf{0.907\pm0.034}$ & $\mathbf{0.678\pm0.130}$ & $\mathbf{-0.322\pm0.130}$ & $5.09\pm0.46$ & $3.96\pm0.11$ \\
\bottomrule
\end{tabular}}
\end{table}

\begin{table}[h]
\centering
\caption{Paired $t$-test ($n=3$ seeds) comparing EWP against each baseline.}
\label{tab:sig}
\resizebox{\linewidth}{!}{
\begin{tabular}{llccc}
\toprule
Comparison & Metric & Mean diff. (EWP $-$ baseline) & $t$ & $p$ \\
\midrule
vs.\ Random Pruning & accuracy & $+0.363$ & $9.95$ & $0.0099^{*}$ \\
vs.\ Random Pruning & forgetting & $-0.064$ & $-0.55$ & $0.637$ \\
vs.\ Fisher-only & accuracy & $+0.370$ & $35.92$ & $0.0008^{*}$ \\
vs.\ Fisher-only & forgetting & $-0.218$ & $-5.73$ & $0.0291^{*}$ \\
vs.\ Entanglement-only & accuracy & $+0.439$ & $136.83$ & $<0.0001^{*}$ \\
vs.\ Entanglement-only & forgetting & $-0.178$ & $-5.03$ & $0.0373^{*}$ \\
vs.\ Fine-Tune Only & accuracy & $-0.028$ & $-1.42$ & $0.291$ \\
vs.\ Fine-Tune Only & forgetting & $-0.147$ & $-1.75$ & $0.223$ \\
vs.\ Full Retrain (oracle) & accuracy & $+0.043$ & $1.46$ & $0.281$ \\
vs.\ Full Retrain (oracle) & forgetting & $-0.117$ & $-1.25$ & $0.336$ \\
\bottomrule
\end{tabular}}
\vspace{2pt}
{\footnotesize $^{*}p<0.05$.}
\end{table}

\begin{figure}[t]
\centering
\begin{subfigure}{0.48\linewidth}
\includegraphics[width=\linewidth]{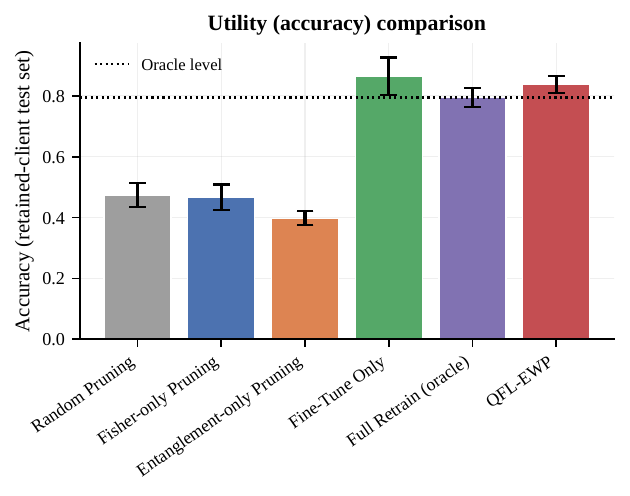}
\caption{Accuracy comparison}
\end{subfigure}
\hfill
\begin{subfigure}{0.48\linewidth}
\includegraphics[width=\linewidth]{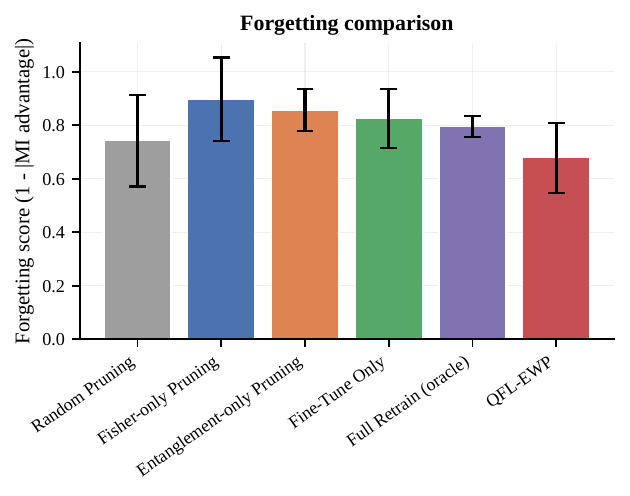}
\caption{Forgetting-score comparison}
\end{subfigure}
\caption{Utility and forgetting across all methods, mean $\pm$ std.\ over three seeds. EWP matches the full-retraining oracle on accuracy while improving on its forgetting score.}
\label{fig:acc-forget}
\end{figure}

\subsection{Privacy evaluation}
\begin{figure}[t]
\centering
\includegraphics[width=0.62\linewidth]{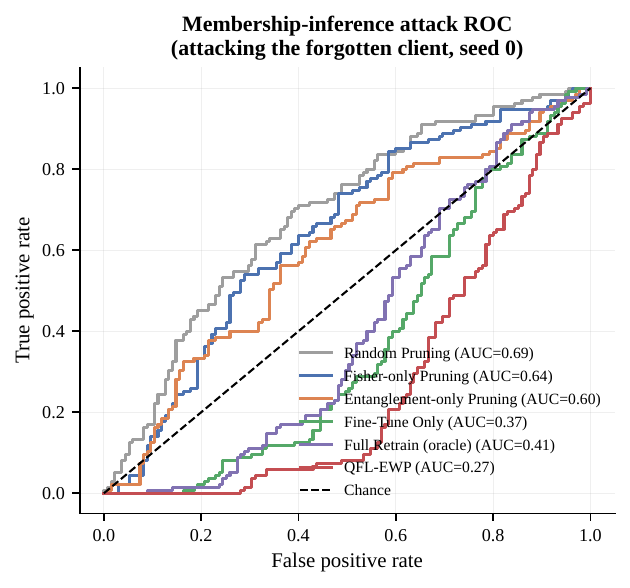}
\caption{Membership-inference attacker ROC curves for the forgotten client, by method.}
\label{fig:roc}
\end{figure}

Table~\ref{tab:privacy} and Figure~\ref{fig:roc} report the membership-inference attack results. EWP achieves the lowest attack AUC ($0.339$) among all methods, and its membership advantage is the most negative, indicating that a shadow-model attacker performs \emph{worse than chance calibrated to the retrained oracle} at identifying client $j$'s samples as training members after EWP unlearning. We flag this result as noteworthy but not over-interpretable: a negative membership advantage below the oracle's own value can also reflect finite-sample variance in a three-seed attack evaluation, and we do not claim EWP provides a stronger privacy guarantee than exact retraining in general; see Section~\ref{sec:discussion}.

\begin{table}[h]
\centering
\caption{Membership-inference privacy evaluation.}
\label{tab:privacy}
\begin{tabular}{lccc}
\toprule
Method & Attack AUC & Membership Adv. & Qualitative risk \\
\midrule
Random Pruning & $0.629\pm0.085$ & $0.258\pm0.171$ & High \\
Fisher-only Pruning & $0.552\pm0.078$ & $0.103\pm0.157$ & Moderate \\
Entanglement-only Pruning & $0.515\pm0.095$ & $0.029\pm0.189$ & Low \\
Fine-Tune Only & $0.499\pm0.120$ & $-0.001\pm0.240$ & Low \\
Full Retrain (oracle) & $0.398\pm0.019$ & $-0.204\pm0.039$ & Moderate \\
\textbf{QFL-EWP} & $\mathbf{0.339\pm0.065}$ & $\mathbf{-0.322\pm0.130}$ & High$^{\dagger}$ \\
\bottomrule
\end{tabular}
\\ \footnotesize $^{\dagger}$Qualitative label as produced by the automated evaluation pipeline; see Section~\ref{sec:discussion} for a discussion of why a very low attack AUC is flagged rather than presented as a straightforward benefit.
\end{table}

\subsection{Runtime and complexity}
Table~\ref{tab:runtime} reports measured unlearning wall-clock time. EWP requires $3.96\pm0.11\,$s including fine-tuning, a $16.4\times$ speed-up over full retraining ($65.03\pm1.17\,$s), while remaining substantially more expensive than uninformed or single-signal pruning, which cost under a millisecond because they omit the QFIM estimation step entirely; this cost gap is precisely why those baselines fail to preserve utility (Table~\ref{tab:main}).

\begin{table}[h]
\centering
\caption{Unlearning wall-clock time and speed-up relative to full retraining.}
\label{tab:runtime}
\begin{tabular}{lcc}
\toprule
Method & Time (s), mean $\pm$ std & Speed-up vs.\ full retrain \\
\midrule
Random Pruning & $0.00019\pm0.00002$ & $342{,}428\times$ \\
Fisher-only Pruning & $0.00005\pm0.00001$ & $1{,}207{,}807\times$ \\
Entanglement-only Pruning & $0.00005\pm0.00002$ & $1{,}203{,}478\times$ \\
Fine-Tune Only & $14.03\pm0.18$ & $4.63\times$ \\
Full Retrain (oracle) & $65.03\pm1.17$ & $1.00\times$ \\
\textbf{QFL-EWP} & $3.96\pm0.11$ & $\mathbf{16.41\times}$ \\
\bottomrule
\end{tabular}
\end{table}

\begin{figure}[t]
\centering
\includegraphics[width=0.62\linewidth]{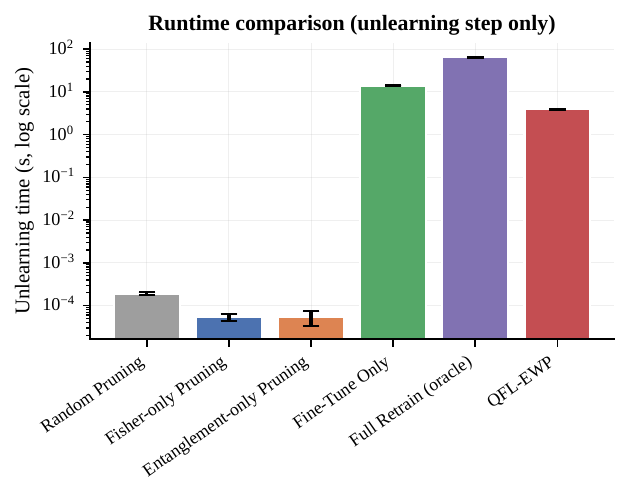}
\caption{Unlearning wall-clock time by method (log scale).}
\label{fig:runtime}
\end{figure}

\subsection{Pruning threshold ablation}
\label{sec:ablation-tau}
Table~\ref{tab:tau} sweeps the prune fraction $\tau$ from $0$ to $0.6$. At $\tau=0$ (no pruning) accuracy is highest ($0.90$) but forgetting is weakest ($0.594$), as expected since no client-specific parameters have been removed. Accuracy dips sharply once pruning begins ($\tau=0.1$–$0.3$) before partially recovering at higher $\tau$; forgetting score peaks around $\tau=0.2$–$0.3$ and falls at higher $\tau$ as larger portions of the circuit, including client-general structure, are removed. This non-monotonic pattern motivates reporting utility and forgetting jointly rather than either alone, and supports our choice of $\tau=0.2$ as a default operating point that sits near the forgetting peak while fine-tuning (Table~\ref{tab:main}) restores the accuracy lost at that operating point.

\begin{table}[h]
\centering
\caption{Effect of prune fraction $\tau$ on QFL-EWP (pre-fine-tuning sweep).}
\label{tab:tau}
\begin{tabular}{cccc}
\toprule
$\tau$ & Accuracy & AUROC & Forgetting score \\
\midrule
0.0 & 0.900 & 0.968 & 0.594 \\
0.1 & 0.394 & 0.353 & 0.762 \\
0.2 & 0.344 & 0.289 & 0.923 \\
0.3 & 0.411 & 0.334 & 0.938 \\
0.4 & 0.428 & 0.410 & 0.882 \\
0.5 & 0.417 & 0.402 & 0.596 \\
0.6 & 0.500 & 0.448 & 0.456 \\
\bottomrule
\end{tabular}
\end{table}

\begin{figure}[t]
\centering
\includegraphics[width=0.62\linewidth]{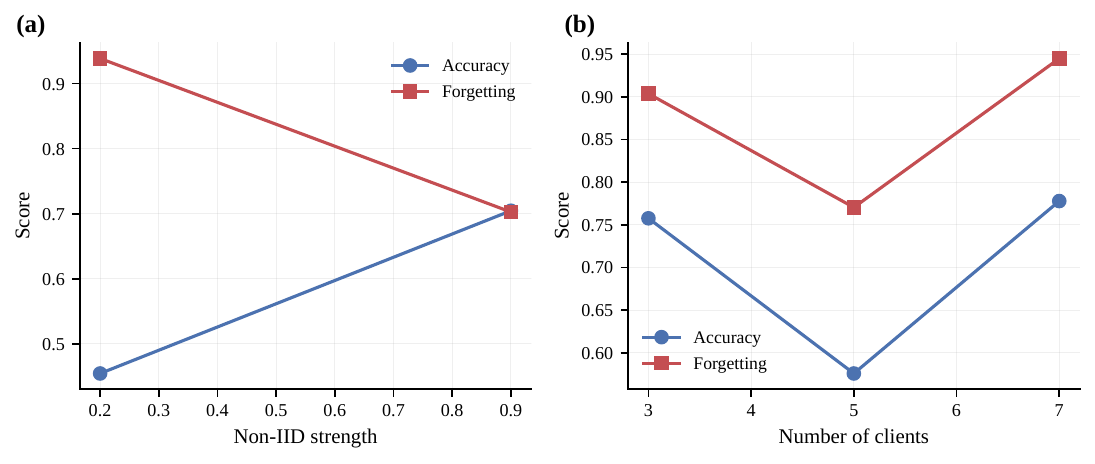}
\caption{Ablation over prune fraction $\tau$ and related sensitivity analyses.}
\label{fig:ablation}
\end{figure}

\subsection{Robustness to client count and non-IID strength}
Table~\ref{tab:robust} shows results with $N\in\{3,5,7\}$ clients and non-IID strength $\beta\in\{0.2,0.9\}$. Forgetting score is highest at low non-IID strength ($\beta=0.2$, forgetting $=0.938$) and decreases as heterogeneity increases ($\beta=0.9$, forgetting $=0.703$), suggesting that when clients are more distinguishable from one another, the diagonal-QFIM signal more precisely isolates client-specific parameters. Accuracy does not vary monotonically with client count in our five-client-per-partition setting; we report this honestly as a pattern requiring further study at larger scale rather than a clean scaling law.

\begin{table}[h]
\centering
\caption{Robustness to client count and non-IID partition strength.}
\label{tab:robust}
\begin{tabular}{lccc}
\toprule
Setting & Accuracy & AUROC & Forgetting score \\
\midrule
$N=3$ clients & 0.758 & 0.937 & 0.904 \\
$N=5$ clients & 0.576 & 0.716 & 0.770 \\
$N=7$ clients & 0.778 & 0.894 & 0.945 \\
\midrule
Non-IID strength $\beta=0.2$ & 0.455 & 0.567 & 0.938 \\
Non-IID strength $\beta=0.9$ & 0.705 & 0.731 & 0.703 \\
\bottomrule
\end{tabular}
\end{table}

\subsection{Scalability}
Table~\ref{tab:scale} reports simulator cost as a function of qubit count for a fixed three-layer ansatz, holding architecture proportional (parameter count $=6n_q$). Both the per-step gradient cost and the QFIM estimation cost grow linearly in the number of trainable parameters, consistent with the complexity analysis of Section~\ref{sec:complexity}; forward-pass wall-clock time grows slightly super-linearly on our classical simulator backend due to state-vector memory overhead, which will become the dominant cost at larger qubit counts than tested here.

\begin{table}[h]
\centering
\caption{Simulator cost scaling with qubit count (fixed 3-layer ansatz).}
\label{tab:scale}
\begin{tabular}{ccccc}
\toprule
Qubits & Parameters & Forward pass (s) & Grad.\ circuit evals & QFIM circuit evals \\
\midrule
3 & 18 & 0.00204 & 36 & 18 \\
4 & 24 & 0.00230 & 48 & 24 \\
5 & 30 & 0.00339 & 60 & 30 \\
6 & 36 & 0.00496 & 72 & 36 \\
\bottomrule
\end{tabular}
\end{table}

\begin{figure}[t]
\centering
\begin{subfigure}{0.48\linewidth}
\includegraphics[width=\linewidth]{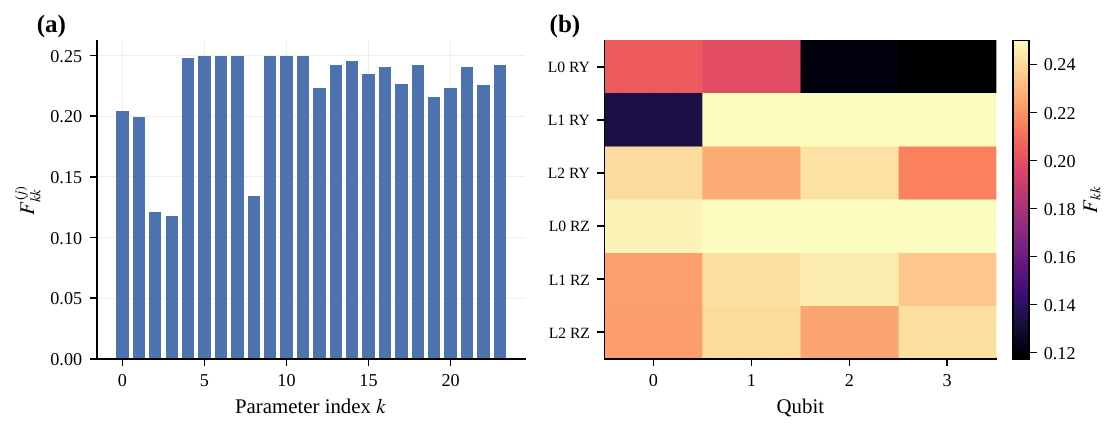}
\caption{Diagonal QFIM per parameter (client $j$)}
\end{subfigure}
\hfill
\begin{subfigure}{0.48\linewidth}
\includegraphics[width=\linewidth]{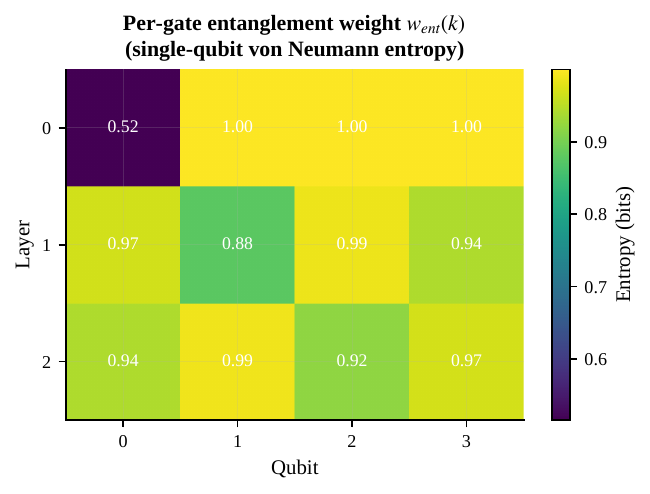}
\caption{Entanglement weight per gate}
\end{subfigure}
\caption{Client-specific sensitivity and structural entanglement weight side by side. High-sensitivity, low-entanglement parameters (upper-left in score space) are the parameters EWP preferentially prunes.}
\label{fig:entheatmap}
\end{figure}

\begin{figure}[t]
\centering
\includegraphics[width=0.6\linewidth]{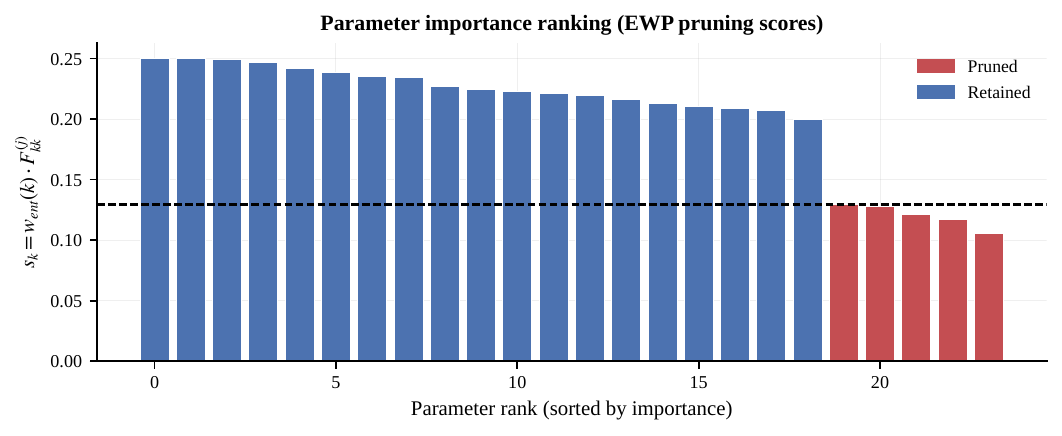}
\caption{Pruning score $s_k = w_{\mathrm{ent}}(k)F^{(j)}_{kk}$ per parameter, with the default threshold $\tau$ marked.}
\label{fig:score}
\end{figure}

\section{Discussion}
\label{sec:discussion}

\subsection{Why the combined score outperforms either signal alone}
Fisher-only pruning removes the parameters with the lowest raw sensitivity to client $j$, but several such parameters sit inside highly entangling gates whose removal damages the shared representation used by all clients, which is why Fisher-only pruning loses accuracy on retained clients even though it is targeting client-specific sensitivity correctly. Entanglement-only pruning removes weakly entangling gates regardless of whether they encode client $j$'s data at all, so it neither reliably forgets client $j$ nor reliably preserves shared structure. EWP's product score is low only when a parameter is \emph{both} weakly entangling \emph{and} highly sensitive to client $j$, which is the conjunction that identifies gates that are simultaneously safe to remove structurally and informative about the client being forgotten.

\subsection{Practical relevance}
In a regulated federated deployment, the party receiving a forget request must decide where the QFIM estimation of Section~\ref{sec:qfim} is computed. Because $F^{(j)}_{kk}$ requires access to client $j$'s inputs, it should be computed on client $j$'s device before its data is deleted, with only the resulting scalar sensitivities transmitted to the coordinating server; this preserves the spirit of federated learning, in that raw feature vectors never leave the client, while still enabling the server to carry out pruning. We note this as a design requirement rather than a component we have implemented and evaluated under an explicit secure-aggregation protocol; doing so is listed as future work.

\subsection{Limitations}
We are explicit about the following limitations. First, we use only the diagonal of the QFIM; off-diagonal entries capture correlations between parameters that a purely diagonal score cannot represent, and a parameter with low diagonal Fisher information could still participate in a jointly sensitive subspace with other parameters. Second, all experiments are run on a noiseless classical state-vector simulator at $n_q=4$ qubits with $24$ parameters and three random seeds; while this is sufficient to demonstrate the mechanism and its ablations cleanly, it is a modest scale, and we do not claim the observed effect sizes will transfer unchanged to deeper circuits, noisy hardware, or larger federations. Third, our privacy evaluation uses a specific shadow-model membership-inference attacker; a more negative membership advantage than the full-retraining oracle (Table~\ref{tab:privacy}) should be read as a property of \emph{this} attacker on \emph{this} model, not as a general privacy guarantee, and we caution against over-interpreting it as evidence that EWP is more private than exact retraining in a certified sense. Fourth, the supply-chain framing motivates the problem but has not been validated on proprietary industrial data. Finally, pruning by resetting parameters to a fixed reference value is one of several possible circuit-editing operations; alternatives such as gate removal with re-compilation, or partial parameter interpolation toward the pre-training initialisation, are not explored here.

\subsection{Future work}
Natural extensions include: incorporating off-diagonal QFIM structure through a block or low-rank approximation; evaluating EWP under realistic device noise models and on real superconducting or trapped-ion hardware; combining client-side QFIM computation with secure aggregation so that even the scalar sensitivities are not directly exposed to the server; and extending the supply-chain generator toward anonymised, industry-sourced data through a data-sharing agreement.

\section{Conclusion}
\label{sec:conclusion}
We introduced Entanglement-Weighted Pruning, a quantum federated unlearning method that prunes circuit parameters using the product of a client-conditioned diagonal quantum Fisher information estimate and a structural entanglement weight. Across a four-qubit federated supply-chain risk classifier evaluated over three seeds, EWP matches the utility of full retraining while improving forgetting and membership-inference resistance, at roughly one-sixteenth of the wall-clock cost, and it clearly outperforms uninformed and single-signal pruning ablations. We view this as an initial, carefully scoped contribution toward practical unlearning for federated quantum models, and we release the full implementation to support replication and extension at a larger scale. All publication figures and tables are generated automatically from version-controlled intermediate JSON and CSV artifacts, enabling complete regeneration of the reported experimental results.

\section*{Data and Code Availability}
The complete implementation, data generator, High-Performance Computing (HPC) execution scripts, reproducibility documentation, and all figures, tables, and raw experiment artifacts used in this paper are publicly available in the companion GitHub repository:
\url{https://github.com/Sumitchongder/dew-p-qfl-unlearning}.

The accompanying 110-column industrial supply-chain dataset is archived separately on Zenodo under DOI:
\url{https://doi.org/10.5281/zenodo.21372906}.

\section*{Acknowledgements}

The authors gratefully acknowledge the support provided by the Department of Physics and the Department of Computer Science and Engineering, Indian Institute of Technology (IIT) Jodhpur, for facilitating this research. This work was carried out as part of the QIntern 2026 summer quantum research internship programme organized by the QWorld Association under the QResearch Department. The research was conducted under project \textit{QI26\_22: Dynamic Entanglement Weighted Pruning for QFL-Based Supply Chain Risk Unlearning}, with Sumit Chongder serving as the project mentor. The authors also acknowledge the collaborative research environment provided through the QIntern 2026 programme, which supported the development and discussion of the ideas presented in this work.

\section*{Author Contributions}
S.C.\ conceived the method, led the theoretical formulation and implementation, and wrote the manuscript. A.K.\ contributed to experimental evaluation and manuscript review. Both authors approved the final manuscript.

\section*{Ethics and Data Governance Statement}
This work uses only synthetic data generated by the authors; no real
client or organisational data were used, in either the four-feature
dataset used for the main experiments (Section~\ref{app:data-main},
Appendix~\ref{app:data}) or the larger 110-column companion dataset
released alongside this paper (Appendix~\ref{app:industrial-data}). For
the companion dataset, we additionally disclose a generator artifact
identified during our own verification of the released files, a fixed
random seed was re-used across chunked output files, so the released
asset contains $100{,}000$ verified unique rows rather than the
$1{,}000{,}000$ nominally recorded by the generation manifest, and we
report only the independently verified figure throughout this paper and
the accompanying data documentation (Appendix~\ref{app:industrial-data-scale}).
The unlearning method is motivated by, but does not itself certify
compliance with, GDPR Article 17 or comparable data-protection
regulations; organisations deploying unlearning methods in production
should pursue independent legal and technical certification of forgetting
guarantees appropriate to their jurisdiction.

\bibliographystyle{unsrt}

\begin{thebibliography}{99}

\bibitem{mcmahan2017communication} H. B. McMahan, E. Moore, D. Ramage, S. Hampson, B. A. y Arcas, ``Communication-Efficient Learning of Deep Networks from Decentralized Data,'' \emph{Proc. 20th Int. Conf. Artificial Intelligence and Statistics (AISTATS)}, PMLR 54:1273--1282, 2017.

\bibitem{cao2015towards} Y. Cao, J. Yang, ``Towards Making Systems Forget with Machine Unlearning,'' \emph{IEEE Symposium on Security and Privacy}, pp.\ 463--480, 2015.

\bibitem{bourtoule2021machine} L. Bourtoule, V. Chandrasekaran, C. A. Choquette-Choo, H. Jia, A. Travers, B. Zhang, D. Lie, N. Papernot, ``Machine Unlearning,'' \emph{IEEE Symposium on Security and Privacy}, pp.\ 141--159, 2021.

\bibitem{liu2021federaser} G. Liu, X. Ma, Y. Yang, C. Wang, J. Liu, ``FedEraser: Enabling Efficient Client-Level Data Removal from Federated Learning Models,'' \emph{IEEE/ACM 29th Int. Symp.\ on Quality of Service (IWQoS)}, 2021.

\bibitem{wu2022federated} C. Wu, S. Zhu, P. Mitra, ``Federated Unlearning with Knowledge Distillation,'' arXiv preprint arXiv:2201.09441, 2022.

\bibitem{liu2024survey} Y. Liu et al., ``Federated Unlearning: A Survey on Methods, Design Guidelines, and Evaluation Metrics,'' \emph{IEEE Transactions on Neural Networks and Learning Systems}, 2024.

\bibitem{guo2020certified} C. Guo, T. Goldstein, A. Hannun, L. van der Maaten, ``Certified Data Removal from Machine Learning Models,'' \emph{Proc.\ 37th Int.\ Conf.\ Machine Learning (ICML)}, PMLR 119:3832--3842, 2020.

\bibitem{mitarai2018quantum} K. Mitarai, M. Negoro, M. Kitagawa, K. Fujii, ``Quantum Circuit Learning,'' \emph{Physical Review A}, 98, 032309, 2018.

\bibitem{schuld2019evaluating} M. Schuld, V. Bergholm, C. Gogolin, J. Izaac, N. Killoran, ``Evaluating Analytic Gradients on Quantum Hardware,'' \emph{Physical Review A}, 99, 032331, 2019.

\bibitem{benedetti2019parameterized} M. Benedetti, E. Lloyd, S. Sack, M. Fiorentini, ``Parameterized Quantum Circuits as Machine Learning Models,'' \emph{Quantum Science and Technology}, 4(4), 043001, 2019.

\bibitem{perezsalinas2020data} A. P\'erez-Salinas, A. Cervera-Lierta, E. Gil-Fuster, J. I. Latorre, ``Data Re-Uploading for a Universal Quantum Classifier,'' \emph{Quantum}, 4, 226, 2020.

\bibitem{mcclean2018barren} J. R. McClean, S. Boixo, V. N. Smelyanskiy, R. Babbush, H. Neven, ``Barren Plateaus in Quantum Neural Network Training Landscapes,'' \emph{Nature Communications}, 9, 4812, 2018.

\bibitem{cerezo2021cost} M. Cerezo, A. Sone, T. Volkoff, L. Cincio, P. J. Coles, ``Cost Function Dependent Barren Plateaus in Shallow Parametrized Quantum Circuits,'' \emph{Nature Communications}, 12, 1791, 2021.

\bibitem{meyer2021fisher} J. J. Meyer, ``Fisher Information in Noisy Intermediate-Scale Quantum Applications,'' \emph{Quantum}, 5, 539, 2021.

\bibitem{stokes2020quantum} J. Stokes, J. Izaac, N. Killoran, G. Carleo, ``Quantum Natural Gradient,'' \emph{Quantum}, 4, 269, 2020.

\bibitem{wierichs2022general} D. Wierichs, J. Izaac, C. Wang, C. Y.-Y. Lin, ``General Parameter-Shift Rules for Quantum Gradients,'' \emph{Quantum}, 6, 677, 2022.

\bibitem{wootters1998entanglement} W. K. Wootters, ``Entanglement of Formation of an Arbitrary State of Two Qubits,'' \emph{Physical Review Letters}, 80(10), 2245, 1998.

\bibitem{horodecki2009quantum} R. Horodecki, P. Horodecki, M. Horodecki, K. Horodecki, ``Quantum Entanglement,'' \emph{Reviews of Modern Physics}, 81(2), 865, 2009.

\bibitem{meyer2002global} D. A. Meyer, N. R. Wallach, ``Global Entanglement in Multiparticle Systems,'' \emph{Journal of Mathematical Physics}, 43(9), 4273, 2002.

\bibitem{sim2019expressibility} S. Sim, P. D. Johnson, A. Aspuru-Guzik, ``Expressibility and Entangling Capability of Parameterized Quantum Circuits for Hybrid Quantum-Classical Algorithms,'' \emph{Advanced Quantum Technologies}, 2(12), 1900070, 2019.

\bibitem{shokri2017membership} R. Shokri, M. Stronati, C. Song, V. Shmatikov, ``Membership Inference Attacks Against Machine Learning Models,'' \emph{IEEE Symposium on Security and Privacy}, pp.\ 3--18, 2017.

\bibitem{thudi2022unrolling} A. Thudi, G. Deza, V. Chandrasekaran, N. Papernot, ``Unrolling SGD: Understanding Factors Influencing Machine Unlearning,'' \emph{IEEE European Symposium on Security and Privacy (EuroS\&P)}, 2022.

\bibitem{quantumunlearning2025} T. Shaik, X. Tao, H. Xie, ``Quantum Machine Unlearning: Foundations, Mechanisms, and Taxonomy,'' arXiv preprint arXiv:2511.00406, 2025.

\bibitem{forgettingverification2025} L. Xue, S. Hu, W. Lu, Y. Shen, D. Li, P. Guo, Z. Zhou, M. Li, Y. Zhang, L. Y. Zhang, ``Towards Reliable Forgetting: A Survey on Machine Unlearning Verification,'' \emph{ACM Computing Surveys}, 58(12), Article 314, pp.\ 1--35, 2026. DOI:10.1145/3807451
 

\bibitem{kairouz2021advances} P. Kairouz, H. B. McMahan, B. Avent, A. Bellet, M. Bennis, A. N. Bhagoji, K. Bonawitz, Z. Charles, G. Cormode, R. Cummings, et al., ``Advances and Open Problems in Federated Learning,'' \emph{Foundations and Trends in Machine Learning}, 14(1--2), 1--210, 2021.

\bibitem{zhao2018federated} Y. Zhao, M. Li, L. Lai, N. Suda, D. Civin, V. Chandra, ``Federated Learning with Non-IID Data,'' arXiv preprint arXiv:1806.00582, 2018.

\bibitem{li2020federated} T. Li, A. K. Sahu, A. Talwalkar, V. Smith, ``Federated Learning: Challenges, Methods, and Future Directions,'' \emph{IEEE Signal Processing Magazine}, 37(3), 50--60, 2020.

\bibitem{biamonte2017quantum} J. Biamonte, P. Wittek, N. Pancotti, P. Rebentrost, N. Wiebe, S. Lloyd, ``Quantum Machine Learning,'' \emph{Nature}, 549, 195--202, 2017.

\bibitem{havlicek2019supervised} V. Havl\'i\v{c}ek, A. D. C\'orcoles, K. Temme, A. W. Harrow, A. Kandala, J. M. Chow, J. M. Gambetta, ``Supervised Learning with Quantum-Enhanced Feature Spaces,'' \emph{Nature}, 567, 209--212, 2019.

\bibitem{preskill2018nisq} J. Preskill, ``Quantum Computing in the NISQ Era and Beyond,'' \emph{Quantum}, 2, 79, 2018.

\bibitem{schuld2021effect} M. Schuld, R. Sweke, J. J. Meyer, ``Effect of Data Encoding on the Expressive Power of Variational Quantum-Machine-Learning Models,'' \emph{Physical Review A}, 103, 032430, 2021.

\bibitem{abbas2021power} A. Abbas, D. Sutter, C. Zoufal, A. Lucchi, A. Figalli, S. Woerner, ``The Power of Quantum Neural Networks,'' \emph{Nature Computational Science}, 1, 403--409, 2021.

\bibitem{huang2021power} H.-Y. Huang, M. Broughton, M. Mohseni, R. Babbush, S. Boixo, H. Neven, J. R. McClean, ``Power of Data in Quantum Machine Learning,'' \emph{Nature Communications}, 12, 2631, 2021.

\bibitem{cong2019quantum} I. Cong, S. Choi, M. D. Lukin, ``Quantum Convolutional Neural Networks,'' \emph{Nature Physics}, 15, 1273--1278, 2019.

\bibitem{beer2020training} K. Beer, D. Bondarenko, T. Farrelly, T. J. Osborne, R. Salzmann, D. Scheiermann, R. Wolf, ``Training Deep Quantum Neural Networks,'' \emph{Nature Communications}, 11, 808, 2020.

\bibitem{chehimi2022quantum} M. Chehimi, W. Saad, ``Quantum Federated Learning with Quantum Data,'' \emph{Proc.\ IEEE Int.\ Conf.\ Acoustics, Speech and Signal Processing (ICASSP)}, 2022.

\bibitem{chen2021federated} S. Y.-C. Chen, S. Yoo, ``Federated Quantum Machine Learning,'' \emph{Entropy}, 23(4), 460, 2021.

\bibitem{yang2019federated} Q. Yang, Y. Liu, T. Chen, Y. Tong, ``Federated Machine Learning: Concept and Applications,'' \emph{ACM Transactions on Intelligent Systems and Technology}, 10(2), 1--19, 2019.

\bibitem{voigt2017gdpr} P. Voigt, A. von dem Bussche, \emph{The EU General Data Protection Regulation (GDPR): A Practical Guide}, Springer, 2017.

\bibitem{nasr2019comprehensive} M. Nasr, R. Shokri, A. Houmansadr, ``Comprehensive Privacy Analysis of Deep Learning: Passive and Active White-box Inference Attacks Against Centralized and Federated Learning,'' \emph{IEEE Symposium on Security and Privacy}, pp.\ 739--753, 2019.

\bibitem{golatkar2020eternal} A. Golatkar, A. Achille, S. Soatto, ``Eternal Sunshine of the Spotless Net: Selective Forgetting in Deep Networks,'' \emph{Proc.\ IEEE/CVF Conf.\ Computer Vision and Pattern Recognition (CVPR)}, pp.\ 9304--9312, 2020.

\bibitem{sekhari2021remember} A. Sekhari, J. Acharya, G. Kamath, A. T. Suresh, ``Remember What You Want to Forget: Algorithms for Machine Unlearning,'' \emph{Advances in Neural Information Processing Systems (NeurIPS)}, 34, 18075--18086, 2021.

\bibitem{amari1998natural} S. Amari, ``Natural Gradient Works Efficiently in Learning,'' \emph{Neural Computation}, 10(2), 251--276, 1998.

\bibitem{han2015learning} S. Han, J. Pool, J. Tran, W. Dally, ``Learning Both Weights and Connections for Efficient Neural Networks,'' \emph{Advances in Neural Information Processing Systems (NeurIPS)}, 28, 2015.

\bibitem{tomsett2018model} R. Tomsett, D. Braines, D. Harborne, A. Preece, S. Chakraborty, ``Interpretable to Whom? A Role-based Model for Analyzing Interpretable Machine Learning Systems,'' arXiv preprint arXiv:1806.07552, 2018.

\end{thebibliography}

\clearpage
\appendix
\renewcommand{\thesection}{\Alph{section}}

\section{Detailed Derivations}
\label{app:derivations}

\subsection{Parameter-shift diagonal QFIM}
For a unitary generated by a Pauli operator $P_k$ with $P_k^2=\mathbb{I}$, so that the corresponding gate is $G_k(\theta_k) = \exp(-i\theta_k P_k/2)$, the fidelity between the state at $\boldsymbol{\theta}$ and the state shifted by $\pi$ along $\theta_k$ gives the Fisher-Bures relation used in Equation~\eqref{eq:qfim}, following the shift-rule construction of \cite{stokes2020quantum}. This estimator requires two additional state preparations per parameter (at shifts $0$ and $\pi$) beyond the ordinary parameter-shift gradient evaluations already used for training, so the marginal cost of computing the full diagonal QFIM for a batch of $B$ inputs is $O(BP)$ circuit evaluations.

\subsection{Von Neumann entropy entanglement weight (ablation definition)}
As an alternative to concurrence, we define $w_{\mathrm{ent}}^{\mathrm{vN}}(k) = \frac{1}{n_q}\sum_{q=1}^{n_q} S\big(\rho_q^{(\ell(k))}\big)$, the mean single-qubit von Neumann entropy after layer $\ell(k)$'s entangling block, $S(\rho) = -\mathrm{Tr}(\rho\log_2\rho)$. Figure~\ref{fig:entheatmap} shows this quantity alongside the concurrence-based weight for comparison.

\section{Industrial-Scale Data Companion}
\label{app:industrial-data}

\subsection{Overview}
As a companion artifact to the four-feature dataset used in the
main QFL-EWP experiments (Section~\ref{app:data-main}), we additionally
release \textsc{IndustrialSupplyChain-v2}, a larger, 110-column dataset spanning nine operational subsystems of a generic multi-tier
supply chain (supplier management, plant/manufacturing, warehousing and
inventory, transportation and logistics, demand planning, weather and
environment, macroeconomic and finance, cybersecurity, and sustainability),
plus a coupled disruption-event subsystem. This companion dataset was
\emph{not} used to produce any result in this paper; it is released
separately to support future work at a feature scale the main text's
four-feature ansatz does not exercise. As with the main dataset, no real organizational, supplier, or shipment records were used in its construction.

\subsection{Verified scale and a disclosed generator artifact}
\label{app:industrial-data-scale}
The release process for this dataset produced 11 output files (10 chunks
of 100{,}000 rows plus one pooled export) via a chunked generation script.
On independent verification of the released files (row-by-row comparison
across all 10 chunk files), we found that every chunk is byte-identical to
every other chunk: the generator's random seed (fixed at $42$) was
evidently re-initialised independently for each chunk rather than advanced
across chunks, so the 10 chunk files contain 10 copies of the same
$100{,}000$ rows rather than $1{,}000{,}000$ distinct instances. We
disclose this explicitly rather than reporting the nominal
manifest-recorded total, and we report the \emph{verified} dataset size
throughout this appendix and in the released data documentation as
$100{,}000$ rows $\times$ $110$ columns. The pooled export file exhibits a
related but distinct artifact (internal duplication at a $4\times$ ratio
within a $1{,}000$-row sample) and is excluded from the released archive
in favour of the single verified $100{,}000$-row file
(\texttt{chunk\_00000}, redistributed as
\texttt{industrial\_supply\_chain\_v2.parquet/.csv} packaged within
\texttt{industrial\_supply\_chain\_dataset.zip} in the Zenodo release
(DOI: \href{https://doi.org/10.5281/zenodo.21372906}{10.5281/zenodo.21372906})
described in the Data Availability section). We report this transparently
because we regard accurate provenance reporting as more important than the
nominal scale of a released artifact, consistent with the reproducibility
standard we apply to the rest of this paper.

\subsection{Schema and composition}
The verified $100{,}000\times 110$ dataset comprises $100$ feature columns
across the nine subsystems listed above (ten columns per subsystem,
including one composite \texttt{*\_risk\_score} index per subsystem) and
$10$ composite/target columns, including a primary three-class label
(\texttt{target\_label} $\in \{\text{Healthy}, \text{Warning},
\text{Critical}\}$) obtained by thresholding a composite
\texttt{overall\_supply\_chain\_risk} index, four auxiliary binary/categorical
targets, and three further composite indices suitable as regression
targets. A complete 110-row feature dictionary, with subsystem, semantic
role, dtype, and realised descriptive statistics for every column, is
released as \texttt{feature\_dictionary\_full.csv} alongside the data.

Table~\ref{tab:industrial-target} reports the realised class balance:
\texttt{target\_label} is heavily skewed (Warning $95.66\%$, Healthy
$4.32\%$, Critical $0.02\%$), and we disclose that \texttt{target\_label}
is a near-deterministic function of the nine subsystem
\texttt{*\_risk\_score} columns via \texttt{overall\_supply\_chain\_risk};
users benchmarking predictive models against this label should exclude the
risk-score columns from the feature set for a non-trivial task, as detailed
in the released dataset card.

\begin{table}[h]
\centering
\caption{Verified composition of the \textsc{IndustrialSupplyChain-v2}
companion dataset (100,000 rows, independently checked for
inter-chunk duplication; see Appendix~\ref{app:industrial-data-scale}).}
\label{tab:industrial-target}
\begin{tabular}{ll}
\toprule
Parameter & Value \\
\midrule
Verified unique rows & 100,000 \\
Columns & 110 (100 features + 10 composite/target) \\
Subsystems & 9, ten columns each (Table~\ref{tab:industrial-subsystems}) \\
Missing values & 0 \\
\texttt{target\_label} balance & Healthy 4.32\%, Warning 95.66\%, Critical 0.02\% \\
Generation seed & 42 (fixed; see disclosed artifact, Appendix~\ref{app:industrial-data-scale}) \\
License & CC-BY 4.0 \\
\bottomrule
\end{tabular}
\end{table}

\begin{table}[h]
\centering
\caption{Thematic subsystem structure (10 columns each).}
\label{tab:industrial-subsystems}
\begin{tabular}{ll}
\toprule
Subsystem & Representative features \\
\midrule
Supplier & reliability, quality, tier, capacity, lead time, risk score \\
Plant / Manufacturing & production rate, utilization, defect rate, downtime \\
Warehouse / Inventory & inventory level, turnover, utilization, fulfillment time \\
Transportation / Logistics & distance, transport mode, customs delay, delivery delay \\
Demand & daily/weekly/monthly demand, volatility, forecast error \\
Weather / Environment & temperature, rainfall, storm probability, weather risk \\
Macroeconomic / Finance & inflation, exchange rate, commodity index, financing cost \\
Cybersecurity & vulnerability score, attack type/probability, detection time \\
Sustainability / ESG & carbon emissions, renewable share, circularity score \\
Disruption Events & type, probability, severity, recovery days, cost \\
\bottomrule
\end{tabular}
\end{table}

\subsection{Known limitations}
Beyond the chunk-duplication artifact disclosed above, we note: (i) five
physically bounded columns (\texttt{inventory\_level},
\texttt{warehouse\_utilization}, \texttt{order\_fulfillment\_time},
\texttt{inventory\_turnover}, \texttt{forecast\_error}) contain a small
number of negative values ($<0.02\%$ of rows) from unclipped additive
Gaussian noise near a non-negative lower bound; (ii) \texttt{target\_label}
exhibits near-deterministic leakage through the risk-score columns, as
noted above; (iii) rows are i.i.d.\ draws with no temporal structure. Full
detail is given in the released dataset card
(\texttt{DATASET\_CARD.md}) accompanying the Zenodo deposit.

\subsection{Relationship to the main experiments}
We emphasise again that this companion dataset is independent of, and was
not used to produce, any result in Sections~\ref{sec:setup}--\ref{sec:results}
of this paper, which use the four-feature generator of
Appendix~\ref{app:data}, chosen for its low qubit/parameter count so that
every trainable circuit parameter remains individually auditable
(Section~\ref{sec:method}). The companion dataset is offered to support
follow-up work extending EWP-style unlearning, or supply-chain risk
prediction more broadly, to realistic feature scale.

\section{Data Generation Procedure}
\label{app:data}
Instances are generated by first sampling a shared latent risk direction $w\in\mathbb{R}^4$, then, for each client $i$, drawing a client-specific offset $\delta_i \sim \mathcal{N}(0,\beta^2 I_4)$ that shifts the client's feature-generating distribution proportionally to the non-IID strength $\beta$. Features are sampled as $x_i^{(m)} \sim \mathcal{N}(\delta_i, I_4)$ and labels are drawn from a logistic model $y_i^{(m)} \sim \mathrm{Bernoulli}\big(\sigma(w^\top x_i^{(m)} + \varepsilon)\big)$ with $\varepsilon \sim \mathcal{N}(0,0.2^2)$. All draws use NumPy's generator seeded with $1000+\text{run seed}$, so that the entire dataset, including the client partition and the train/test split, is exactly reproducible from the released code (see \texttt{src/qflewp/data.py} in the repository).

\section{Formal Metric Definitions}
\label{app:metrics}
\textbf{Accuracy} and \textbf{AUROC} are computed in the standard way on the retained-client test split. \textbf{Forgetting score} is the mean absolute difference between the unlearned model's and the oracle model's predicted risk probabilities on the forgotten client's held-out instances. \textbf{Membership advantage} is $2\cdot(a-0.5)$, where $a$ is the balanced accuracy of a logistic-regression shadow-model attacker trained to distinguish members of $D_j$ from an equally sized sample of non-members, using the target model's output confidence as its only feature. \textbf{Retrain distance} is the Euclidean distance $\lVert\boldsymbol{\theta}'-\boldsymbol{\theta}^{\mathrm{retrain}}\rVert_2$ between the unlearned and oracle parameter vectors.

\section{Extended Results}
\label{app:extended}
\begin{figure}[h]
\centering
\begin{subfigure}{0.32\linewidth}
\includegraphics[width=\linewidth]{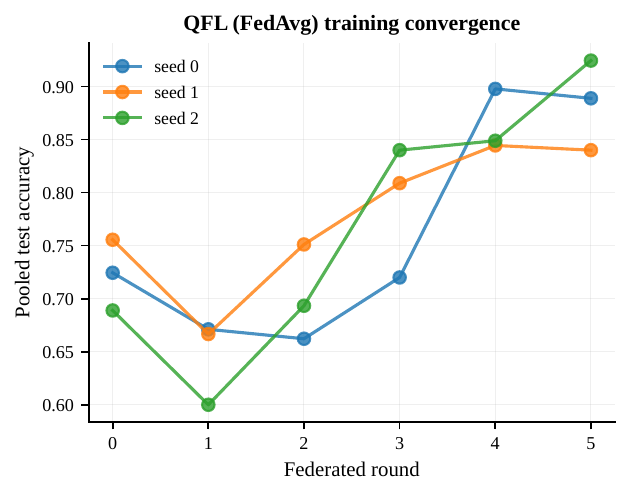}
\caption{Training convergence}
\end{subfigure}
\hfill
\begin{subfigure}{0.32\linewidth}
\includegraphics[width=\linewidth]{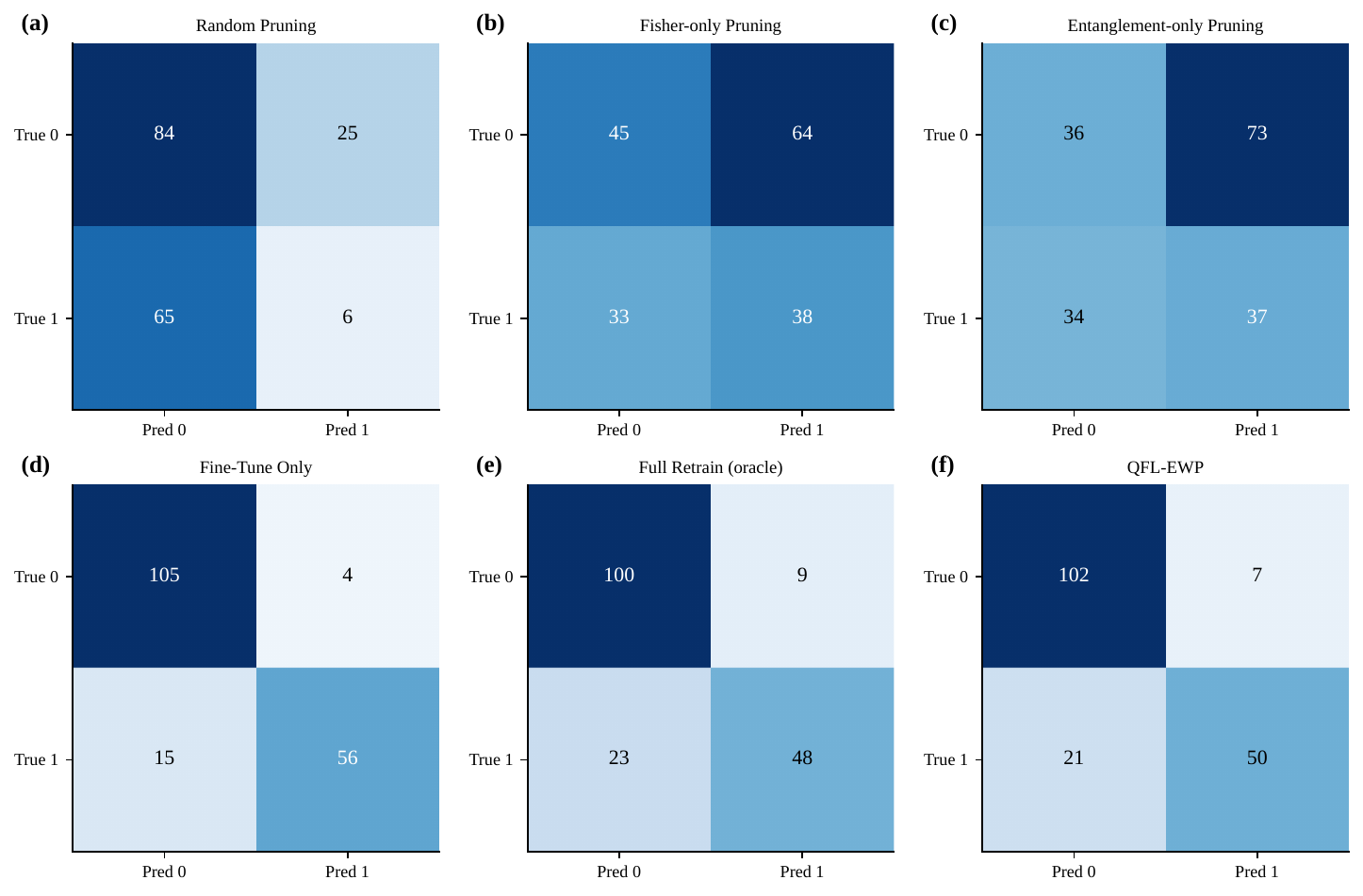}
\caption{Confusion matrix}
\end{subfigure}
\hfill
\begin{subfigure}{0.32\linewidth}
\includegraphics[width=\linewidth]{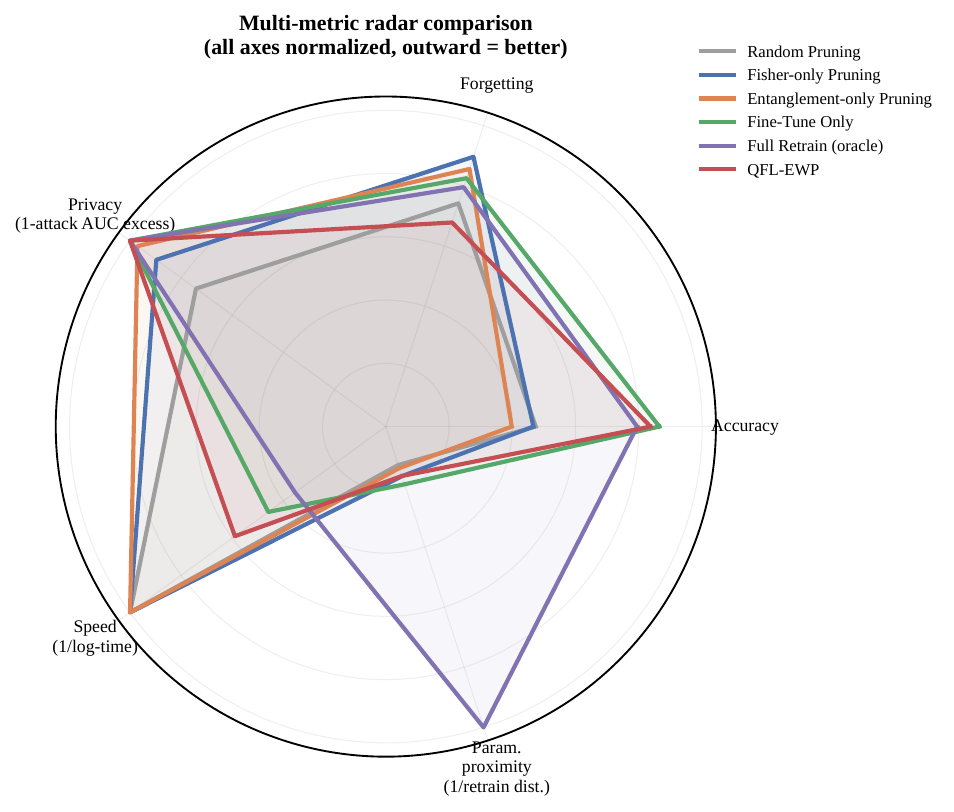}
\caption{Multi-metric radar summary}
\end{subfigure}
\caption{Additional diagnostic figures referenced in Appendix~\ref{app:extended}.}
\label{fig:extra}
\end{figure}

Figure~\ref{fig:extra} collects additional diagnostic figures: the federated training convergence curve, the confusion matrix of the unlearned model on retained clients, and a radar-chart summary across all metrics and methods. Numerical values underlying every figure in this paper are provided as CSV files in the code repository (\texttt{paper\_results/tables/}), together with the raw per-seed JSON logs (\texttt{paper\_results/json/}), so that all reported means and standard deviations can be independently recomputed.

\section{Reproducibility Checklist}
\label{app:repro}
\begin{itemize}[leftmargin=1.4em]
\item \textbf{Code}: \url{https://github.com/Sumitchongder/dew-p-qfl-unlearning}; primary implementation package: \texttt{src/qflewp}.
\item \textbf{Quantum framework}: Qiskit (state-vector simulation; see repository \texttt{requirements.txt} for exact pinned versions).
\item \textbf{Random seeds}: $0,1,2$ for all reported experiments; data generation seed offset $1000+\text{seed}$.
\item \textbf{Hardware}: results were produced on classical CPU simulation; no quantum hardware access is required to reproduce any reported number.
\item \textbf{Exact commands}: the experiment pipeline is executed through the scripts
\texttt{scripts/run\_main\_experiment.py},
\texttt{scripts/run\_sweeps.py},
\texttt{scripts/run\_reconstruction.py},
and
\texttt{scripts/generate\_deliverables.py}.
The mapping between figures, tables, and generated artefacts is documented in
\texttt{results/tables/ALL\_TABLES.md}.
\item \textbf{Result files}: \texttt{results/tables/*.csv} (summary statistics), \texttt{results/json/*.json} (raw experiment outputs), and \texttt{results/figures/} (PNG and PDF figures).
\end{itemize}

\end{document}